\documentclass[aps,nofootinbib,prd,eqsecnum,showpacs,showkeys,preprintnumbers, referee]{revtex4-2}
\usepackage{mathrsfs}
\usepackage{yfonts}
\usepackage{tipa}
\usepackage{bm}
\usepackage{bbm}
\usepackage{amsmath}
\usepackage{amssymb}
\usepackage{amsthm}
\usepackage{amsfonts}
\usepackage{mathtools}
\usepackage{latexsym}
\usepackage{relsize}
\usepackage[english]{babel}
\usepackage{booktabs}
\usepackage[utf8]{inputenc}
\usepackage{tikz}
\usepackage{array}
\usepackage{hyperref, color}
\usepackage{tikz}
\usetikzlibrary{shapes.geometric,arrows.meta,positioning,shapes.symbols}

\newcommand{\bea}{\begin{eqnarray}}
\newcommand{\eea}{\end{eqnarray}}
\newcommand{\be}{\begin{equation}}
\newcommand{\ee}{\end{equation}}
\newcommand{\ba}{\begin{align}}
\newcommand{\ea}{\end{align}}

\begin{document}
\date{\today}
\title{Cosmological test of a length-preserving biconnection gravity}
\author{D. Mhamdi$^{1}$}
\email{dalale.mhamdi@ump.ac.ma}
\author{A. Bouali$^{1,2}$}
\email{a1.bouali@ump.ac.ma}
\author{T. Ouali$^{1}$}
\email{t.ouali@ump.ac.ma}
\affiliation{\textsuperscript{1}Laboratory of Physics of Matter and Radiations, Mohammed I University, BP 717, Oujda, Morocco,\\  \textsuperscript{2}Higher School of Education and Training, Mohammed I University, BP 717, Oujda, Morocco}
\author{T. Harko$^{3,4}$}
\email{tiberiu.harko@aira.astro.ro}
\affiliation{${}^3$Department of Physics, Babe\c s-Bolyai University, Kog$\breve{a}$lniceanu Street, Cluj-Napoca 400084, Romania,}
\affiliation{${}^4$Astronomical Observatory, 19 Cire\c silor Street, Cluj-Napoca 400487, Romania.}
\date{\today }
\begin{abstract} 
We investigate the cosmological implications of an extended gravitational framework based on biconnection gravity, constructed from the Schr$\ddot{o}$dinger connection and its dual. In this approach, the difference between the two connections defines the mutual curvature, which encodes the non-Riemannian geometric degrees of freedom, while their symmetric combination reduces to the Levi-Civita connection and hence reproduces general relativity at the background level. Within this setting, we derive the generalized Friedmann equations for a spatially flat Friedmann-Lema\^{i}tre-Robertson-Walker Universe. The resulting  equations contain additional geometric contributions that may naturally encode an effective dark energy sector induced by the biconnection  degrees of freedom.  We explore this extra dark energy by adopting  five commonly used parametrizations, namely B$\Lambda$CDM, $\omega$CDM, Chevallier-Polarski-Linder, Barboza-Alcaniz, and a logarithmic equations of state. These considerations are confronted with recent observational data, including DESI DR2, Pantheon$^+$, and CC observations. Our analysis shows that the four parameterizations enter the acceleration phase at almost the same redshifts and share the same current value of the Hubble rate.
Furthermore, the statistical comparison based on the Akaike, Bayesian, and Deviance Information Criterion shows that Barboza-Alcaniz, and  logarithmic parameterizations have strong evidence and are competitive with $\Lambda$CDM. To classify this biconnection gravity  in the plethora theoretical models describing the current cosmic acceleration, we examine its implications  through cosmographic tools, including the deceleration, jerk, and snap
parameters, as well as through the Statefinder analysis and $Om(z)$ diagnostic. These diagnostics indicate that the geometric
sector generates a dynamical dark energy component that remains observationally close to $\Lambda$CDM at the background
level. Indeed, the effective evolution of the biconnection gravity begins initially from a quintessence-like regime and subsequently evolves toward a phantom-like regime. All parameterizations under consideration face this transition in the recent past, except for the $\omega$CDM model, which enters the phantom-like regime in the future.
\end{abstract}

	\maketitle
          \tableofcontents
          
\section{Introduction}
One of the most important achievements in cosmology today is the observational evidence that the Universe is currently undergoing an accelerated expansion. Within the framework of General Relativity (GR), this phenomenon is usually explained by introducing dark energy, an exotic component with negative pressure. The simplest and most accepted description of dark energy is provided by the $\Lambda$CDM (Lambda Cold Dark Matter) model, in which dark energy is identified with a cosmological constant, $\Lambda$, characterized by an equation of state parameter equal to -1. Despite its remarkable success in fitting current cosmological data, the $\Lambda$CDM model suffers from several theoretical shortcomings, including the fine-tuning \cite{w1989,s2000} and the coincidence \cite{si2013,ve2014} problems, respectively. These issues have motivated extensive efforts to construct viable alternatives to the cosmological constant. Effective scalar field theories are such viable construction, including quintessence \cite{fu1982,fo1987,we1988}, phantom models \cite{bo2015,bou2021,mh2023,dah20,dahmani2023}, and k-essence \cite{ch2000,malq2003}, as well as holographic dark energy models \cite{hor2000,li2004,wan2017,be2012,ba2021,bo2011}, relativistic fluid approaches such as Chaplygin gas \cite{gor2003}, and dynamical dark energy scenarios, notably running vacuum energy models \cite{geor2018,reza2019,sola2017}. For the present status of the $\Lambda$CDM cosmology see \cite{Him1} and \cite{Him2}, respectively.

From another perspective, the challenge of explaining the late-time accelerated expansion of the Universe has motivated the exploration of a variety of gravitational theories beyond GR. In this context, a broad class of modified gravity models has been introduced and extensively studied. Indeed, to explain the available cosmological observations beyond the standard $\Lambda$CDM framework which is based on  Riemannian geometry, several extensions based on non-Riemannian framework have been suggested. Among these alternatives, the $f(R)$ extension to  the Hilbert-Einstein \cite{buc1970}, $f(R,L_m)$ theory \cite{harko2010}, $f(R, T)$ gravities \cite{harko2011},  and the Hybrid Metric Palatini theory \cite{harko2012}. For more details on the extensions of $f(R)$ gravity, see \cite{harko2019}.  Other extensions to Riemannian geometry like Finsler and Weyl geometries,  were studied in \cite{hama2023,hama2022,boua2023} and \cite{ghilencea2023,ghilen2023,conde2024,harko2024}, respectively. The elegant construction of Weyl geometry faces a severe issue as the length of a vector is not conserved through an autoparallel transport.\\

In this context, and in order to overcome the length variability problem, specific to the standard formulation of Weyl geometry,  Schr$\ddot{o}$dinger proposed an intriguing class of affine connections \cite{sch1985}. Schr$\ddot{o}$dinger's connections provide a minimal and torsion-free generalization of the Levi-Civita connection, in which nonmetricity is introduced in a controlled manner. Although such connections possess nonmetricity, they preserve the length of vectors under autoparallel transport \cite{min2024}.  The implications of a gravitational action constructed from a length preserving non-metricity, in the absence of torsion, were considered in \cite{min2024}. The corresponding theory was investigated in both Palatini and metric formalisms. While the Palatini variation leads to standard general relativity, the metric version of the theory contributes with several nonmetricity dependent extra terms in the gravitational Einstein field equations,which can be interpreted as representing a geometric type of dark energy. The cosmological investigations of the Weyl-Schr\"{o}dinger theory, as performed in \cite{min2024}, did show that it represents an attractive, and viable alternative to standard general relativity, in which dark energy can be explained as a purely geometric effect. 

An important mathematical concept, developed initially in the framework of the geometric theory of the statistical manifolds \cite{statman1, statman2,statman3}, is the concept of dual connection, which can be naturally defined on a statistical manifold. In the framework of the gravitational physics, biconnection gravitational theories have been investigated in \cite{bi1,bi2,bi3,bi4,bi5}. The physical and cosmological implications of the Schr$\ddot{o}$dinger connections have been investigated in Refs.~\cite{klem2021,iosi2023,csil2024}. 

Based on the mathematical concept of dual  connection, a new biconnection gravity theory,  has been constructed in \cite{,Csillag:2024bvc} by using the Schr$\ddot{o}$dinger connection and its dual, which also carries nonmetricity with an opposite sign. Interestingly enough, the symmetric combination of the aforementioned connections reduces identically to the Levi-Civita connection. As a result, the gravitational field equations evaluated on a homogeneous and isotropic background coincide with those of GR. The dynamic of this construction is encoded in the mutual tensor, defined as the difference between the two affine connections i.e. the Schr$\ddot{o}$dinger connection and its dual. This tensor collects the independent non-Riemannian degrees of freedom of the biconnection framework. Its contributions modify the field equations through additional terms in the effective Einstein equations, thereby inducing departures from the standard cosmological dynamics. 

In this paper, we test and assess the biconnection gravity, developed in details in \cite{Csillag:2024bvc}, with respect the most recent datasets namely DESI DR2, Pantheon$^+$, and CC. 
To achieve this goal, we  first review the biconnection theory, constructed from the well known mutual curvature scalar by adopting a metric approach rather than the metric-affine framework considered in \cite{iosi2023}. Furthermore, the  two connections are chosen to be  the Schr$\ddot{o}$dinger connection and its dual,  without introducing any direct coupling between geometry and matter. We assume that the field equations of the biconnection theory retain the same structural form as those of the Einstein's theory, with the key distinction that the Ricci tensor and Ricci scalar are replaced by the mutual curvature tensor and mutual curvature scalar, respectively. The corresponding Friedmann equations for a homogeneous, isotropic, and spatially flat Universe are derived. The system of cosmological equations is totally determined by imposing suitable constraints on the model parameters. Furthermore, we parametrize the extra terms appearing in the Friedmann equations by five equations of state. Indeed, we consider B$\Lambda$CDM, $\omega$CDM parameterizations as well as Chevalier-Polarski-Linder, Barboza-Alcaniz and logarithmic  parameterizations.

To constrain the cosmological parameters of this length preserved biconnection gravity, we perform a Markov Chain Monte Carlo (MCMC) analysis \cite{pad2021} utilizing a combination of recent observational datasets. These include the Baryon Acoustic Oscillation (BAO) data (including the latest results DR2 from the Dark Energy Spectroscopic Instrument (DESI)), Pantheon+ compilation of Type Ia supernovae, and cosmic chronometer measurements. To assess the performance of our model, we establish a comparison  to the standard $\Lambda$CDM scenario using the Akaike Information Criterion \cite{lidd2007,akai1974}, the Bayesian Information Criterion \cite{schw1978}, and the Deviance Information Criterion \cite{spieg2002}. Furthermore, we explore how these constraints affect the current accelerated expansion of the Universe by analyzing the evolution of key cosmographic quantities: the deceleration parameter q, the jerk parameter j, and the snap parameter s. Since these parameters are constructed from successive time derivatives of the scale factor a(t), they offer a model-independent approach to characterizing cosmic dynamics. Additionally, to enable a geometric comparison between different cosmological scenarios, we adopt the statefinder and $Om(z)$ diagnostics. The latter is particularly useful for distinguishing between different dark energy models or, more generally, between different effective dark energy fluids. \\

This paper is organized as follows. In Sec.~\ref{sec1}, we provides an overview of the biconnection gravity framework and summarize the main theoretical ingredients relevant for our analysis. In Sec.~\ref{sec2}, we derive the generalized Friedmann equations for a Friedmann-Lema\^{i}tre-Robertson-Walker spacetime. Sec.~\ref{sec3} is devoted to the parameterization of the effective dark energy sector, namely the B$\Lambda$CDM, $\omega$CDM, Chevallier-Polarski-Linder, Barboza-Alcaniz, and logarithmic parameterizations. In Sec.~\ref{sec4}, we describe the observational datasets and the methodology adopted to constrain the cosmological parameters. The aforementioned parameterizations are further investigated through cosmographic analysis, the statefinder  and $Om(z)$ diagnostics in Sec.~\ref{sec5}. Finally, Sec.~\ref{sec6} summarizes our main findings and presents our conclusions.

\section{An overview of biconnection gravity}\label{sec1}

In the present Section we briefly review the theoretical foundations of the biconnection gravity theory as introduced in \cite{Csillag:2024bvc}. The generalized Einstein and the cosmological evolution equations are also written down.  
\subsection{Mutual curvature}
To formulate the generalized Einstein equations within a non-Riemannian geometric framework and investigate their fundamental properties, we adopt the geometric formalism based on the Schr$\ddot{o}$dinger connection, and its dual. The Schr$\ddot{o}$dinger connection provides a simple and nontrivial extension of Riemannian geometry through the presence of nonmetricity. Its dual connection, characterized by an opposite nonmetricity, naturally complements this geometric structure. The biconnection gravity is equipped with these two connections and form what is known as statistical structure on manifolds.

The symmetric combination of the Schr$\ddot{o}$dinger connection and its dual reduces to the Levi-Civita connection, thereby recovering general relativity at the background level. In contrast, their difference encodes additional geometric degrees of freedom responsible for deviations from Riemannian geometry. This difference defines the so-called mutual tensor, which plays a central role in the biconnection framework. The associated mutual curvature tensor, $\mathcal{R}^\lambda_{\;\;\rho\mu\nu}$, is constructed from the Schr$\ddot{o}$dinger curvature tensor $R^\lambda_{\;\;\rho\mu\nu}$, its dual $\tilde{R}^\lambda_{\;\;\rho\mu\nu}$, and the mutual tensor $\mathcal{K}^\lambda_{\;\;\sigma\mu}$, as detailed in Ref.~\cite{Csillag:2024bvc}.
 \begin{eqnarray}
\mathcal{R}^\lambda_{\;\;\;\rho\mu\nu}&=&\frac{1}{2}\left({R}^\lambda_{\;\;\;\rho\mu\nu}+{\tilde{R}}^\lambda_{\;\;\;\rho\mu\nu}\right)\\
&-&\frac{1}{2}\mathcal{K}^\lambda_{\;\;\sigma\mu}\mathcal{K}^\sigma_{\;\;\rho\nu}+\frac{1}{2}\mathcal{K}^\lambda_{\;\;\sigma\nu}\mathcal{K}^\sigma_{\;\;\rho\mu},\nonumber
\end{eqnarray}
where the expression of the mutual tensor, $\mathcal{K}^\lambda_{\;\;\sigma\mu}$, is given by
\begin{equation}
\mathcal{K}^\lambda_{\;\;\sigma\mu}=\Gamma^\lambda_{\;\;\sigma\mu}-\tilde{\Gamma}^{\lambda}_{\;\;\sigma\mu},
\end{equation}
with $\Gamma^\lambda_{\;\;\sigma\mu}$ and $\tilde{\Gamma}^{\lambda}_{\;\;\sigma\mu}$ denote the Schr$\ddot{o}$dinger connection and its dual, respectively.
  Henceforth, 'tilde' will refer to quantities related to the dual of the Schr$\ddot{o}$dinger connection. \\
  
The mutual Ricci tensor and the mutual scalar are given by
\begin{eqnarray}\label{MR}
\mathcal{R}_{\rho\nu}&=&\frac{1}{2}\left({R}_{\rho\nu}+{\tilde{R}}_{\rho\nu}\right)\\
&-&\frac{1}{2}\mathcal{K}^\lambda_{\;\;\sigma\lambda}\mathcal{K}^{\sigma}_{\;\;\;\;\rho\nu}+\frac{1}{2}\mathcal{K}^{\lambda}_{\;\;\sigma\nu}\mathcal{K}^\sigma_{\;\;\rho\lambda},\nonumber
\end{eqnarray}	
\begin{eqnarray}\label{MS}
\mathcal{R}&=&\frac{1}{2}\left({R}+{\tilde{R}}\right)
-\frac{1}{2}\mathcal{K}^\lambda_{\;\;\sigma\lambda}\mathcal{K}^{\sigma\;\;\;\;\rho}_{\;\;\;\;\rho}+\frac{1}{2}\mathcal{K}^{\lambda\;\;\;\;\rho}_{\;\;\sigma}\mathcal{K}^\sigma_{\;\;\rho\lambda},
\end{eqnarray}	
respectively.\\

To obtain the analogue of the Einstein equations by means of the mutual geometries,  we first recall that a metric affine geometry is characterized by a metric tensor, $g_{\mu\nu}$, and a connection given by
\begin{eqnarray}
\Gamma^\mu_{\;\;\nu\rho}&=&\hat{\Gamma}^\mu_{\;\;\nu\rho}+\frac{1}{2}g^{\lambda\mu}\left(-Q_{\lambda\nu\rho}+Q_{\rho\lambda\nu}+Q_{\nu\rho\lambda}\right)\nonumber\\
&-&\frac{1}{2}g^{\lambda\mu}\left(-T_{\lambda\rho\nu}+T_{\rho\nu\lambda}+T_{\nu\rho\lambda}\right)
\end{eqnarray}
where $\hat{\Gamma}^\mu_{\;\;\nu\rho}$, $Q_{\lambda\nu\rho}$ and $T_{\lambda\rho\nu}$ are the Levi-Civita connection, non metricity and torsion, respectively. From now quantities with a 'hat' refer to the Levi-Civita connection. In this paper, we consider  the free torsion Schr$\ddot{o}$dinger connection characterized by the following nonmetricity by means of the one form $\pi_\mu$ \cite{Csillag:2024bvc}
\begin{eqnarray}
Q_{\mu\nu\rho}&=&\pi_\mu g_{\nu\rho}-\frac{1}{2}\left(\pi_\rho g_{\mu\nu}+\pi_\nu g_{\mu\rho}\right).
\end{eqnarray}	
A straightforward calculations give the Ricci tensor and the scalar curvature 
\begin{eqnarray}
{R}_{\mu\nu}&=&{\hat{R}}_{\mu\nu}-g_{\mu\nu}\hat{\nabla}_\alpha \pi^\alpha+\frac{1}{2}\hat{\nabla}_\mu \pi_\nu-\hat{\nabla}_\nu \pi_\mu\nonumber\\
&-&g_{\mu\nu}\pi^\alpha \pi_\alpha  -\frac{1}{4}\pi_\mu \pi_\nu,
\end{eqnarray}
\begin{eqnarray}
{R}&=&{\hat{R}}-\frac{9}{2}\hat{\nabla}_\mu \pi^\mu
 -\frac{9}{4}\pi_\mu \pi^\mu,
\end{eqnarray}	
respectively.

Furthermore, the other pillar required to construct the biconnection gravity is the dual of the Schr$\ddot{o}$dinger connection. The dual of the Schr$\ddot{o}$dinger connection is  characterized by the following nonmetricity and  torsion 
\begin{eqnarray}
\tilde{Q}_{\mu\nu\rho}&=&-\pi_\mu g_{\nu\rho}+\frac{1}{2}\left(\pi_\rho g_{\mu\nu}+\pi_\nu g_{\mu\rho}\right)\\
\tilde{T}_{\rho\mu\nu}&=&\frac{3}{2}\left(\pi_\nu g_{\mu\rho}-\pi_\mu g_{\nu\rho}\right),\nonumber
\end{eqnarray}
respectively.	
The corresponding  Ricci tensor and scalar curvature constructed from the dual of the Schr$\ddot{o}$dinger connection write	
\begin{eqnarray}
{\tilde{R}}_{\mu\nu}&=&{\hat{R}}_{\mu\nu}-\frac{1}{2}g_{\mu\nu}\hat{\nabla}_\alpha \pi^\alpha+\hat{\nabla}_\mu \pi_\nu+\hat{\nabla}_\nu \pi_\mu\nonumber\\
&+&g_{\mu\nu}\pi^\alpha \pi_\alpha  +\frac{1}{2}\pi_\mu \pi_\nu,
\end{eqnarray}
\begin{eqnarray}
{\tilde{R}}&=&{\hat{R}}+\frac{9}{2}\pi^\mu \pi_\mu,
\end{eqnarray}
respectively.\\

Hence, from Eqs. (\ref{MR}) and (\ref{MS}), the mutual Ricci tensor and the scalar curvature  which are the pillar of a gravitational theory based on two connections write
\begin{eqnarray}\label{MR2}
{\mathcal{R}}_{\mu\nu}&=&{\hat{R}}_{\mu\nu}-\frac{3}{4}g_{\mu\nu}\hat{\nabla}_\alpha \pi^\alpha+\frac{3}{4}\hat{\nabla}_\mu \pi_\nu\nonumber\\
&-&\frac{1}{4}g_{\mu\nu}\pi^\alpha \pi_\alpha  -\frac{1}{8}\pi_\mu \pi_\nu,
\end{eqnarray}	
\begin{eqnarray}\label{MS2}
{\mathcal{R}}&=&{\hat{R}}-\frac{9}{4}\hat{\nabla}_\mu \pi^\mu
 -\frac{9}{8}\pi_\mu \pi^\mu,
\end{eqnarray}	
respectively.
\subsection{Einstein's equations}
To write the analogue of the Einstein's equations, we assume that the field equations of the biconnection theory preserve the structural form of Einstein’s equations, with the Ricci tensor and scalar curvature  replaced by the mutual curvature tensor and mutual scalar curvature, respectively. Furthermore, the two connections are fixed to the Schr$\ddot{o}$dinger connection and its dual. Hence,  the gravitational field equations take the form \cite{Csillag:2024bvc}
\begin{eqnarray}\label{MEE}
{\mathcal{R}}_{(\mu\nu)}-\frac{1}{2}g_{\mu\nu}{\mathcal{R}}=8\pi{T}_{\mu\nu}.
\end{eqnarray}	

The above consideration are motivated by the fact that the antisymmetric contributions, ${\mathcal{R}}_{[\mu\nu]}$, to the Einstein equations are commonly associated with non standard matter sources and usually emerge in frameworks involving  geometry-matter coupling. In this study, we restrict ourselves to the metric formalism and exclude any geometry-matter coupling, with particular emphasis on cosmological applications. Furthermore, in a Friedmann-Lema\^itre-Robertson-Walker geometry, the vector $\pi^\mu$ admits only a temporal component, and therefore the antisymmetric part, i.e. $\mathcal{R_{[\mu\nu]}}$, vanishes identically.\\

Taking into account Eqs. (\ref{MR2}) and (\ref{MS2}), the Einstein's equations, Eqs (\ref{MEE}), write
\begin{eqnarray}\label{GGF}
{\hat{R}}_{\mu\nu}&-&\frac{1}{2}g_{\mu\nu}\left({\hat{R}}+\frac{3}{4}\hat{\nabla}_\alpha \pi^\alpha+\frac{5}{8}\pi^\alpha \pi_\alpha \right) +\frac{3}{8}\hat{\nabla}_\mu \pi_\nu+\frac{3}{8}\hat{\nabla}_\nu \pi_\mu\nonumber\\
&+& -\frac{1}{8}\pi_\mu \pi_\nu=8\pi{T}_{\mu\nu}.
\end{eqnarray}	
\section{Friedmann equations}\label{sec2}

In this section, we examine the cosmological implications of the biconnection gravity given by the generalized gravitational field, Eq. (\ref{GGF}). To this aim, we consider a spatially flat Friedmann-Lema\^itre-Robertson-Walker spacetime
\begin{equation}
	ds^2=-dt^2+a^2(t)\delta_{ij}dx^idx^j,
\end{equation}
and the Universe  filled with a perfect fluid described by the following energy momentum tensor 
\begin{equation}\label{fluid}
	T_{\mu\nu}=(\rho+p)u_\mu u_\nu+pg_{\mu\nu}.
\end{equation} 
where the four velocity, $u_\mu$, is normalized such that 
$u^\mu u_\mu=-1$, $\rho$ and $p$ denote the energy density and pressure of the fluid, respectively.  Furthermore, as the Universe is homogeneous and isotropic, the temporal component of  the field $\pi$ depends only on time \cite{Csillag:2024bvc}
\begin{equation}\label{pi}
	\pi^\sigma=(\phi(t),0,0,0).
\end{equation}
Taking into account the above considerations, the corresponding generalized Friedmann equations are expressed as 
\begin{equation}\label{BF1}
	3H^2=8\pi\rho+\frac{9}{8}\dot{\phi}+\frac{9}{8}H\phi-\frac{3}{16}
\phi^2=8\pi(\rho +\rho_{\text{eff}})
\end{equation}
\begin{equation}\label{BF2}
	2\dot{H}+3H^2=-8\pi p+\frac{3}{8}\dot{\phi}+\frac{15}{8}H\phi-\frac{5}{16}
\phi^2=-8\pi(p +p_{\text{eff}})
\end{equation}
where we have introduced the effective energy density 
\begin{equation}
	\rho_{\text{eff}}=\frac{1}{8\pi}\left(\frac{9}{8}\dot{\phi}+\frac{9}{8}H\phi-\frac{3}{16}
\phi^2\right),
\end{equation}
and the effective pressure
\begin{equation}
	p_{\text{eff}}=-\frac{1}{8\pi}\left(\frac{3}{8}\dot{\phi}+\frac{15}{8}H\phi-\frac{5}{16}
\phi^2\right).
\end{equation}
The equation of conservation are given from Eqs. (\ref{BF1}) and (\ref{BF2}) as
\begin{equation}
	(\dot{\rho}+\dot{\rho}_{\text{eff}})+3H(\rho+\rho_{\text{eff}}+p+p_{\text{eff}})=0
\end{equation}

In order to constrain observationally our theoretical model, we rewrite the biconnection Fridemann equations with respect to the redshift variable, $z$, using the following time redshift relation 
\begin{equation}
	\frac{d}{dt}=-(1+z)H(z)\frac{d}{dz}.
\end{equation}

Consequently, the dimensionless  Friedmann equations take the form
\begin{eqnarray}\label{NF1}
h^2(z)&=&r(z)-\frac{3}{8}(1+z)h(z)\frac{d\Phi(z)}{dz}+\frac{3}{8}h(z)\Phi(z)-\frac{1}{16}\Phi^2(z)\label{NF1}\\
-2(1+z)h(z)\frac{dh(z)}{dz} +3h^2(z)&=&-P
-\frac{3}{8}(1+z)h(z)\frac{d\Phi(z)}{dz}+\frac{15}{8}h(z)\Phi(z)-\frac{5}{16}\Phi^2(z)\label{NF2}
\end{eqnarray}

where we have used the following set of dimensionless variables 
\begin{equation}
H=H_0h,\quad \rho=\rho_cr,\quad p=\frac{1}{3}\rho_cP,\quad \phi=H_0\Phi,
\end{equation}
with $H_0$ and $\rho_c=3H_0/8\pi$ denote the current value of the Hubble rate and the critical energy density, respectively.
\section{Dark energy equations of state}\label{sec3}
To close the above system of equations, we parametrize that the effective energy density and pressure of the length preserving biconnection gravity by the equation of state (EoS)
\begin{equation}
p_{\text{eff}}=\omega_{de}(z)\rho_{\text{eff}},
\end{equation}
where $\omega(z)$ is the dynamical EoS parameter characterizing the effective dark energy density. With this choice, the system of equation is closed and it is solved by imposing the initial condition $h(z=0)=1$ and $\Phi(z=0)=\Phi_0$ as a free parameter. The remaining parameter, the matter energy density is derived from Eq. (\ref{NF1}).\\

In this paper, we parameterize the biconnection gravity by considering five form of the dynamical EoS $\omega(z)$. The particular case, $\omega = -1$, representing the cosmological constant in the context of the biconnection gravity which we label the $\mathrm{B}\Lambda\mathrm{CDM}$ model. The second EoS, labeled by $\omega$CDM, is characterized by a free constant parameter, $\omega$.   The remaining  three parameterizations of the dark energy equation of state are Chevallier-Polarsky-Linder (CPL) \cite{linder}
\begin{equation}
\omega_{de}(z)=\omega_0+\omega_1\frac{z}{1+z},
\end{equation}
Barboza-Alcaniz \cite{Barboza}
\begin{equation}
\omega_{de}(z)=\omega_0+\omega_1\frac{z(1+z)}{1+z^2},
\end{equation}
and logarithmic form \cite{log}
\begin{equation}
\omega_{de}(z)=\omega_0+\omega_1\ln(1+z).
\end{equation}
\section{Cosmological constraints}\label{sec4}
In this section, we examine the observational viability of the length preserving biconnection gravity model using cosmological data.  In order to achieve this goal, we parametrize the theory through three equations of state, namely the CPL, Barboza-Alcaniz, and logarithmic parametrizations, in addition to the $\omega$CDM and $\mathrm{B}\Lambda\mathrm{CDM}$ models. The viability of the model is assessed through a statistical comparison between the theoretical predictions of length-preserving biconnection gravity and cosmological observations. More specifically, we perform a statistical inference of the parameter set $(\Phi_0, \omega_0, \omega_a, h)$ for each equation of state parametrization using the Markov Chain Monte Carlo (MCMC) technique~\cite{pad2021}. Here, the dimensionless Hubble parameter $h$ is related to the present day Hubble constant via $H_0 = 100\,h\,\mathrm{km\,s^{-1}\,Mpc^{-1}}$. Our analysis incorporates the following observational datasets: 
the Pantheon$^+$ Type Ia supernova compilation~\cite{pantheon}, the cosmic chronometer (CC) data~\cite{cc,cc1,cc2,cc3,cc4} and 
the latest baryon acoustic oscillation (BAO) measurements from the Dark Energy Spectroscopic Instrument (DESI DR2) \cite{DESI1}. 
\subsection{Datasets}
Baryon Acoustic Oscillations (BAO), arising from sound waves in the early universe, serve as a standard ruler for measuring cosmological distances and enable precise tests of cosmology. BAO measurements constrain the comoving distance $D_M(z)$, the Hubble distance $D_H(z)$, and the volume averaged distance $D_V(z)$, defined  as
\begin{equation}
D_M(z) \equiv \int_{0}^{z} \frac{c\,dz'}{H(z')},
\end{equation}
\begin{equation}
D_H(z) \equiv \frac{c}{H(z)},
\end{equation}
and
\begin{equation}
D_V(z)=\left[z D_M(z)^2 D_H(z)\right]^{1/3},
\end{equation}
respectively. These distance measures are expressed relative to the sound horizon scale at the drag epoch, $r_d$,  defined as
\begin{equation}
r_d \equiv r_s(z_d)=\int_{z_d}^{\infty}\frac{c_s(z')}{H(z')}\,dz',
\label{rs}
\end{equation}
where $z_d$ denotes the redshift of the drag epoch and $c_s$ is the sound speed of the baryon--photon fluid.
The Dark Energy Spectroscopic Instrument (DESI) Data Release 2 (DR2) \cite{DESI2} provides high-precision BAO measurements over the redshift range $0.295<z<2.33$, significantly improving upon DESI DR1 \cite{DESI1} with increased statistical power. DESI DR2 provides the following observables:
\begin{equation}
\left(\frac{D_M(z)}{r_d}\right)_{\rm obs}
= \frac{(1+z)D_A(z)}{r_d},
\qquad
\left(\frac{D_H(z)}{r_d}\right)_{\rm obs}
= \frac{c}{H(z)r_d},
\quad\text{and}\quad
\left(\frac{D_V(z)}{r_d}\right)_{\rm obs}
= \frac{\left[z D_M(z)^2 D_H(z)\right]^{1/3}}{r_d}.
\label{BAO_obs}
\end{equation}
We use the BAO distance measurements $D_M/r_d$, $D_H/r_d$, and $D_V/r_d$ from DESI DR2, as reported in Table II of~\cite{DESI2}. To investigate the cosmological implications of the biconnection EoS models using DESI DR2 data, we further include the following complementary datasets:
\begin{itemize}
    \item \textbf{Type Ia Supernovae (SNe Ia):} We combine the DESI data with the Pantheon$^+$ compilation, which consists of 1701 light curves corresponding to 1550 spectroscopically confirmed supernovae in the redshift range $10^{-3} < z < 2.27$. The dataset provides the observed distance modulus $\mu^{\mathrm{obs}}$ at each redshift $z$~\cite{pantheon}.
    
    \item \textbf{Cosmic Chronometers (CC):} We include 36 $H(z)$ measurements obtained using the differential age method~\cite{cc,cc1,cc2,cc3,cc4}.

\end{itemize}
Throughout this paper, we perform our analysis using the combined DESI DR2 + Pantheon$^+$ + CC dataset.

\subsection{Methodology and information criteria}
A consistent comparison between theoretical models and observational data 
requires a statistically robust inference framework. Given the Bayesian 
nature of cosmology, model parameters are generally inferred within a 
Bayesian statistical approach,
\begin{equation}
\mathcal{P}(\boldsymbol{\theta} \mid D) 
\propto 
\mathcal{L}(D \mid \boldsymbol{\theta}) ,
\end{equation}
where $\mathcal{L}(D \mid \boldsymbol{\theta})$ denotes the likelihood function 
and $\mathcal{P}(\boldsymbol{\theta} \mid D)$ the posterior distribution. 
Here, $D$ represents the full observational dataset, while 
$\boldsymbol{\theta}$ denotes the vector of cosmological parameters. The specific parameter vector depends on the model under consideration. 
For the $\Lambda$CDM model and $B\Lambda$CDM, the vector is
$\boldsymbol{\theta}_{\Lambda\mathrm{CDM}}=(\Omega_m, h)$ and $\boldsymbol{\theta}_{B\Lambda\mathrm{CDM}} = (\Phi_0, h)$, respectively. In the $w$CDM parametrization the parameter vector becomes 
$\boldsymbol{\theta}_{w\mathrm{CDM}} = (\Phi_0, \omega_0, h)$. Finally, for the CPL, Barboza-Alcaniz, and logarithmic dark-energy parametrizations, 
an additional free parameter $\omega_a$ is introduced, such that $\boldsymbol{\theta}_{\mathrm{DE}} = (\Phi_0, \omega_0, \omega_a, h)$.
In the case of Gaussian distribution of data, the chi-square function, $\chi^{2}(\boldsymbol{\theta})$, is related to the likelihood function, $ \mathcal{L}(\boldsymbol{\theta})$, by
\begin{equation}
    \chi^2(\boldsymbol{\theta}) = -2 \ln \mathcal{L}(\boldsymbol{\theta}).
\end{equation}
To assess the support of length-preserving biconnection gravity from observational data, 
we perform a Markov Chain Monte Carlo (MCMC) analysis. The cosmological parameters 
are constrained by minimizing the total chi-square function, $\chi^2_{\mathrm{tot}}$, 
constructed from the combined datasets\footnote{For uncorrelated data points, the chi-square statistic is defined as 
$\chi^2 = \sum_{i=1}^{N} \frac{(D_i - T_i)^2}{\sigma_i^2}$, 
where $D_i$ denotes the observed value, $T_i$ the theoretical prediction, 
and $\sigma_i$ the associated uncertainty. 
In the presence of correlations, the chi-square generalizes to the matrix form 
$\chi^2 = \Delta \mathbf{D}^{\mathrm{T}} \mathbf{C}^{-1} \Delta \mathbf{D}$, 
where $\Delta \mathbf{D} = \mathbf{D} - \mathbf{T}$ is the residual vector and 
$\mathbf{C}$ is the covariance matrix including both statistical and systematic uncertainties.}, 
\begin{align}
\chi^2_{\mathrm{tot}} 
=  \chi^2_{\mathrm{DESI}} +\chi^2_{\mathrm{Pantheon}^{+}}+\chi^2_{CC}.
\end{align}
where $\chi^2_{\mathrm{DESI}}$, $\chi^2_{\mathrm{Pantheon}^{+}}$ and  $\chi^2_{CC}$ are the chi-square of DESI BAO, Pantheon+ and CC data. To compare competing models and determine which provides the best description 
of the observational data, we employ standard statistical information criteria. 
In particular, We consider the corrected Akaike Information Criterion (AIC$_c$) 
\cite{akai1974,AIC2}, the Bayesian Information Criterion (BIC) \cite{lidd2007,schw1978}, 
and the Deviance Information Criterion (DIC) \cite{spieg2002,lidd2007}, 
defined respectively as
\begin{equation}
AIC_c = \chi^2_{\min} + 2K_f 
+ \frac{2K_f (K_f + 1)}{N_t - K_f - 1},
\end{equation}
\begin{equation}
\mathrm{BIC} = \chi^2_{\min} + K_f \ln N_t,
\end{equation}
\begin{equation}
\mathrm{DIC} = \bar{D} + p_D,
\end{equation}
where $\chi^2_{\min}$ is the minimum of the chi-square, $K_f$ is the number of free parameters and $N_t$ is the total number of 
data points. In the DIC definition, 
$\bar{D} = \overline{D(\boldsymbol{p})}$ denotes the posterior mean of the 
Bayesian deviance, with $D(\boldsymbol{p}) = \chi^2_{\mathrm{tot}}(\boldsymbol{p})$, 
and $p_D = \bar{D} - D(\bar{\boldsymbol{p}})$ is the effective number of parameters. 
The overline indicates an average over the posterior distribution.\\
For a given criterion, the model with the smallest value is statistically preferred. 
Following common practice in the literature, we adopt the $\Lambda$CDM model 
as the reference model.
To quantify the relative performance of alternative models with respect to 
the reference, we compute
\begin{equation}
\Delta AIC_c = AIC_{c,\mathrm{model}} - AIC_{c,\Lambda\mathrm{CDM}},
\end{equation}
\begin{equation}
\Delta BIC = BIC_{\mathrm{model}} - BIC_{\Lambda\mathrm{CDM}},
\end{equation}
\begin{equation}
\Delta DIC = DIC_{\mathrm{model}} - DIC_{\Lambda\mathrm{CDM}}.
\end{equation}
The strength of evidence 
against a model relative to the reference is commonly interpreted as follows\footnote{$\Delta X$ denotes $\Delta AIC_c$, $\Delta BIC$, and $\Delta DIC$.}: 
$|\Delta X|$ $< 2$ indicates statistically comparable support, 
$2 \le |\Delta X| < 4$ weak evidence against the model, 
$4 \le |\Delta X| < 6$ positive evidence against  the model, 
$6 \le |\Delta X| < 10$ strong evidence against the model
and $|\Delta X| \ge 10$ decisive evidence against the model.
\section{Results and discussions}\label{sec5}

In the present Section we present the statistical results obtained for the considered biconnection model, and we analyze their cosmological implications.  
\begin{widetext}
\begin{table}[t]
\centering
\begin{tabular}{lcccccc}
\hline\hline
Parameters & $\Lambda$CDM & B$\Lambda$CDM & $w$CDM & CPL & Barboza-Alcaniz & Logarithmic \\
\hline\hline
\multicolumn{7}{c}{\textbf{DESI DR2 + Pantheon$^+$ + CC}} \\

$\Phi_0$ 
& - 
& $1.139 \pm 0.021$
& $1.117 \pm 0.029$ 
& $1.118 \pm 0.030$ 
& $1.119 \pm 0.029$ 
& $1.119 \pm 0.030$ \\

$w_0$ 
& -
& -
& $-1.071^{+0.071}_{-0.062}$ 
& $-1.04^{+0.16}_{-0.14}$ 
& $-1.04 \pm 0.12$ 
& $-1.03 \pm 0.13$ \\

$w_a$ 
& -
& -
& -
& $-0.24 \pm 0.71$ 
& $-0.17^{+0.45}_{-0.34}$ 
& $-0.26^{+0.60}_{-0.51}$ \\

$\Omega_m$ 
& $0.2911 \pm 0.0083$ 
& -
& -
& -
& -
& - \\

$h_0$ 
& $0.7110 \pm 0.0079$ 
& $0.7092 \pm 0.0075$
& $0.7071 \pm 0.0082$ 
& $0.7071 \pm 0.0079$ 
& $0.7072 \pm 0.0084$ 
& $0.7077 \pm 0.0082$ \\

$r_d$ 
& $143.5 \pm 1.6$ 
& $143.0 \pm 1.4$
& $142.5 \pm 1.6$ 
& $142.5 \pm 1.4$ 
& $142.5 \pm 1.6$ 
& $142.4 \pm 1.5$ \\

$M$ 
& $-19.330 \pm 0.023$ 
& $-19.330 \pm 0.022$
& $-19.330 \pm 0.023$ 
& $-19.332 \pm 0.021$ 
& $-19.331 \pm 0.023$ 
& $-19.330 \pm 0.023$ \\
\hline
$\chi^2_{\min}$ 
& 1575.302
& 1570.585
& 1569.547
& 1569.563 
& 1569.648 
& 1569.619 \\
\hline
\multicolumn{7}{c}{\textbf{Derivative parameters}} \\

$r_0$
& - 
& $0.31^{+0.0096}_{-0.0096}$
& $0.38^{+0.06}_{-0.06}$ 
& $0.33^{+0.15}_{-0.15}$
& $0.33^{+0.13}_{-0.12}$
& $0.33^{+0.15}_{-0.15}$ \\

$q_0$
& $-0.564^{+0.012}_{-0.012}$
& $-0.538^{+0.014}_{-0.014}$
& $-0.49^{+0.04}_{-0.04}$ 
& $-0.52^{+0.08}_{-0.08}$
& $-0.52^{+0.07}_{-0.07}$
& $-0.52^{+0.08}_{-0.08}$ \\

$z_t$
& $0.695^{+0.023}_{-0.023}$
& $0.705^{+0.024}_{-0.024}$
& $0.73 \pm 0.03$ 
& $0.72 \pm 0.05$
& $0.72^{+0.05}_{-0.05}$
& $0.73^{+0.05}_{-0.06}$ \\

$j_0$
& $1.00$ 
& $0.22^{+0.02}_{-0.02}$
& $0.16^{+0.06}_{-0.06}$ 
& $0.47^{+0.62}_{-0.57}$
& $0.40^{+0.35}_{-0.38}$
& $0.47^{+0.62}_{-0.58}$ \\

$s_0$
& $-0.309^{+0.037}_{-0.037}$
& $-2.87^{+0.031}_{-0.031}$
& $-2.21^{+0.59}_{-0.60}$ 
& $-5.84^{+5.78}_{-6.33}$
& $-4.32^{+3.18}_{-2.72}$
& $-5.75^{+5.58}_{-6.10}$ \\
\hline\hline
\end{tabular}
\caption{Marginalized cosmological constraints (68\% C.L.) for the 
$\Lambda$CDM, B$\Lambda$CDM, $w$CDM, CPL, Barboza-Alcaniz, and logarithmic 
EoS parametrizations obtained from the joint DESI DR2 + Pantheon$^{+}$ + CC dataset. 
The table also reports the derived cosmographic parameters: the present matter 
density parameter $r_0$, the deceleration parameter $q_0$, the transition 
redshift $z_t$, the jerk parameter $j_0$, and the snap parameter $s_0$, 
together with the minimum chi-square values $\chi^2_{\min}$ for each model.}
\label{mcmc_models}
\end{table}
\end{widetext}

\begin{widetext}
\begin{figure}[t!]
\includegraphics[width=0.8\textwidth]{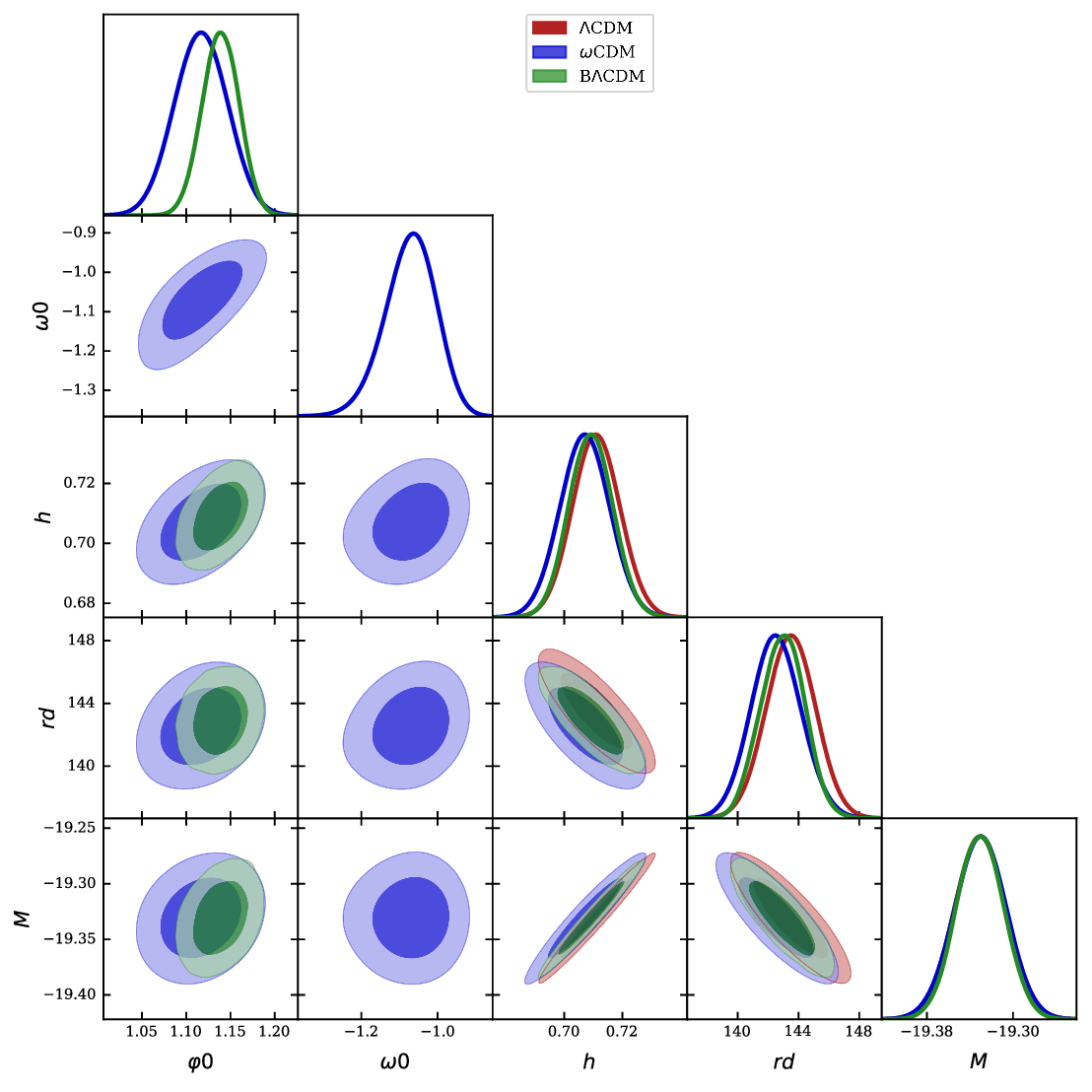}
\caption{The $1\sigma$ and $2\sigma$ confidence contours obtained from the 
combined DESI DR2 + Pantheon$^{+}$ + CC datasets for $\Lambda$CDM (red), 
$w$CDM (blue), and B$\Lambda$CDM (green).}
\label{triangle1}
\end{figure}
\end{widetext}
\subsection{Statistical Results}
\begin{widetext}
\begin{figure}[t!]
\includegraphics[width=0.8\textwidth]{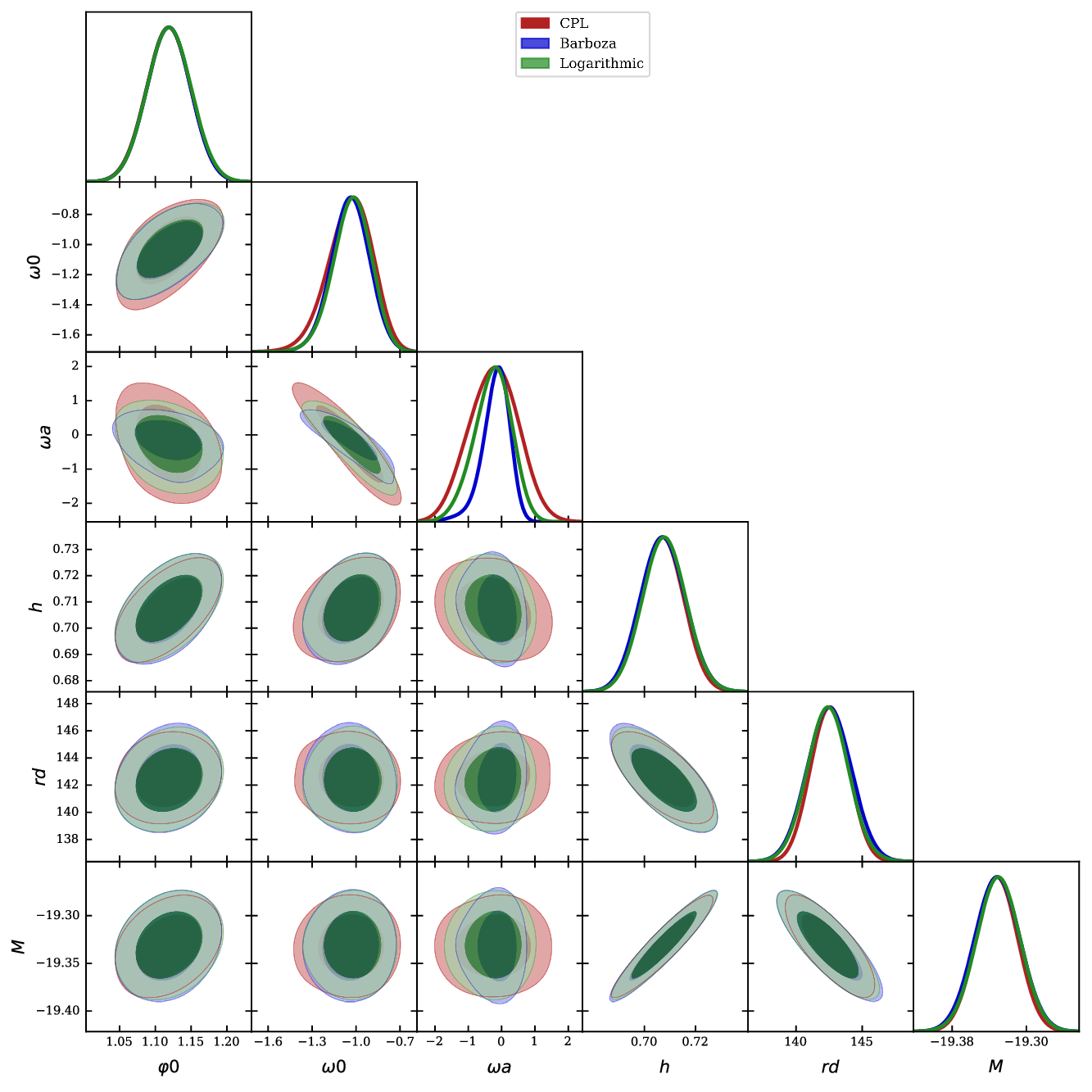}
\caption{The $1\sigma$ and $2\sigma$ confidence contours obtained from the combined DESI DR2 + Pantheon$^{+}$ + CC dataset datasets for the CPL (red), Barboza-Alcaniz (blue), and logarithmic (green) parametrizations.}\label{triangle2}
\end{figure}
\end{widetext}
In Table \ref{mcmc_models}, the mean $\pm 1\sigma$ of the cosmological parameters of equations of state and of the biconnection gravity are shown considering the combination DESI DR2 + Pantheon$^+$ + CC datasets. For each parameterizations, the minimum of the chi-square as well as $\Delta AIC_c$, $\Delta BIC$ and $\Delta DIC$ are shown in the table \ref{tab:AIC_BIC_DIC_refLCDM_all}.  

\paragraph{Confidence contours.} The 2D marginalized confidence contours and 1D posterior distributions of this data combination  are shown in Fig. \ref{triangle1} for $\Lambda$CDM, $w$CDM, and B$\Lambda$CDM while those of CPL, Barboza-Alcaniz, and logarithmic parameterizations are shown in Fig. \ref{triangle2}. We notice that Figs.~\ref{triangle1} and \ref{triangle2} show the same behavior between parameters. For instance, $\Phi_0$ and $\omega_0$ show positive correlation for all parameterizations, no clear correlation between $\Phi_0$ and $\omega_a$ while negative correlation is shown for $\omega_0$ and $\omega_a$.

\paragraph{Hubble parameter values.} From Table \ref{mcmc_models}, we get 
$H_0 = 71.1\pm 0.79\;{\rm km\;s}^{-1}\; {\rm Mpc}^{-1}$, $H_0 = 70.71\pm 0.82\;{\rm km.s}^{-1}\; {\rm Mpc}^{-1}$, $H_0 = 70.92 \pm 0.75 \;\mathrm{km\,s^{-1}\,{\rm Mpc}^{-1}}$, $H_0 = 70.71\pm 0.79\;{\rm km.s}^{-1}\;{\rm  Mpc}^{-1}$, $H_0 = 70.72\pm 0.84\;{\rm km.s}^{-1}\; {\rm Mpc}^{-1}$, and $H_0 = 70.77\pm 0.82\;{\rm km\;s}^{-1}\; {\rm Mpc}^{-1}$ for $\Lambda$CDM, $w$CDM, B$\Lambda$CDM, CPL, Barboza-Alcaniz, and logarithmic, respectively. This indicates that the description of the biconnection gravity by these parameterizations does not affect the current value of the Hubble rate. Indeed, except the $\Lambda$CDM model, we notice that all equations of state share almost the same current value of the Hubble rate.  

\paragraph{The Hubble tension.} To quantify the tension of these values with the  SH0ES value,   we calculate their deviation to $H_0^{\mathrm{SH0ES}} = 73.2 \pm 1.3\;\mathrm{km\,s^{-1}\,Mpc^{-1}}$ \cite{77}. The tension is at $1.38\sigma$, $1.62\sigma$, $1.5\sigma$, $1.64\sigma$, $1.60\sigma$, and $1.58\sigma$ for the $\Lambda$CDM, $w$CDM, B$\Lambda$CDM, CPL, Barboza-Alcaniz, and logarithmic parameterizations, respectively. These are a promising results regarding the Hubble tension issue. 

However, before a definitive conclusion, more data are mandatory at this level specially the Planck one.  From this Table, we observe that the torsion parameter of the length-preserving biconnection gravity model, $\Phi_0$, takes nearly the same value, $\Phi_0 \simeq 1.11$, for the $\omega$CDM, CPL, Barboza--Alcaniz, and Logarithmic parameterizations. Moreover, when fixing $w_0 = -1$ in the $B\Lambda$CDM model, the best-fit value increases to $\phi_0 \simeq 1.14$. This shift corresponds to a tension of approximately $0.7\,\sigma$, indicating full statistical consistency between the models. All considered parameterizations yield values of $w_0$ statistically compatible with $-1$. 

The largest deviation, found in $\omega$CDM, corresponds to $1.0\,\sigma$, while CPL, Barboza--Alcaniz, and logarithmic models deviate by only $0.25\,\sigma$, $0.33\,\sigma$, and $0.23\,\sigma$, respectively. These results provide no statistically significant indication of deviations from $\Lambda$CDM. Turning to the dynamical component, the parameter $w_a$ remains compatible with zero at well below the $1\sigma$ level in all models. Although its central values are negative, their statistical insignificance does not provide evidence for a redshift evolution of the dark energy equation of state. 

Similarly, the sound horizon at the drag epoch, $r_d$, remains stable across all models, with variations fully consistent within the statistical uncertainties. Likewise, the supernova absolute magnitude $M$ shows negligible shifts, indicating that the biconnection framework leads to nearly the same late-time cosmological predictions as general relativity, with only a marginal impact on supernova calibration and early-universe physics.\\
\begin{table}[t!]
\centering
\caption{Statistical comparison of cosmological models using AIC, BIC, and DIC for the combined DESI DR2 + Pantheon$^{+}$ + CC dataset. The $\Lambda$CDM model is taken as the reference.}
\label{tab:AIC_BIC_DIC_refLCDM_all}
\begin{tabular}{lccccccccc}
\hline\hline
Model & $\chi^2_{\min}$ & $K_f$ & $\chi^2_{\mathrm{d.o.f}}$ 
& AIC$_c$ & BIC & DIC 
& $\Delta$AIC$_c$ & $\Delta$BIC & $\Delta$DIC \\
\hline
$\Lambda$CDM 
& 1575.302 & 4 & 0.903 
& 1583.302 & 1605.170 & 1584.022 
& 0 & 0 & 0 \\

B$\Lambda$CDM
& 1570.585 & 4 & 0.901 
& 1578.585 & 1600.445 & 1578.926 
& $-4.717$ & $-4.725$ & $-5.096$ \\

$\omega$CDM 
& 1569.547 & 5 & 0.900 
& 1579.547 & 1606.882 & 1580.721 
& $-3.755$ & $+1.712$ & $-3.301$ \\

CPL 
& 1569.563 & 6 & 0.900 
& 1581.563 & 1614.365 & 1581.844 
& $-1.739$ & $+9.195$ & $-2.178$ \\

Barboza-Alcaniz 
& 1569.648 & 6 & 0.900 
& 1581.648 & 1614.450 & 1582.274 
& $-1.654$ & $+9.280$ & $-1.749$ \\

Logarithmic 
& 1569.619 & 6 & 0.900 
& 1581.619 & 1614.421 & 1582.106 
& $-1.683$ & $+9.251$ & $-1.916$ \\
\hline\hline
\end{tabular}
\end{table}

\paragraph{Statistical comparison with $\Lambda$CDM.} In Table~\ref{tab:AIC_BIC_DIC_refLCDM_all}, we report the values of 
$\chi^2_{\mathrm{d.o.f}}$\footnote{We define the reduced chi-square as 
$\chi^2_{\mathrm{d.o.f}} = \chi^2_{\min}/(N_t - K_f)$.}, $AIC_c$, $BIC$, $DIC$, and their corresponding differences relative to $\Lambda$CDM. 
We first observe that all biconnection parameterizations, including B$\Lambda$CDM, lead to nearly identical values of $\chi^2_{\mathrm{d.o.f}}\simeq 0.9$. This indicates that, at the level of goodness of fit, the models perform comparably and cannot be distinguished solely through the reduced chi-square statistic. 

Therefore, a meaningful comparison requires the use of model selection criteria that account for both the fit quality and the number of free parameters. When ranking the models according to the minimum values of the Akaike information criterion and the deviance information criterion, we find the same ordering in both cases. B$\Lambda$CDM emerges as the preferred model, followed by $\omega$CDM, CPL, logarithmic, and Barboza--Alcaniz parameterizations, while $\Lambda$CDM appears as the least favored model for the dataset considered. This ordering indicates that, under AIC$_c$ and DIC, the improvement in fit provided by the biconnection framework compensates for the increase in model complexity. In contrast, the Bayesian information criterion, which imposes a stronger penalty on additional parameters, yields a slightly different ranking. 

Although B$\Lambda$CDM remains the most favored model, $\Lambda$CDM moves to second place, followed by $\omega$CDM, while CPL, logarithmic, and Barboza--Alcaniz parameterizations become less competitive. This shift reflects the stronger complexity penalization inherent in BIC, which naturally advantages models with fewer free parameters.

Using the estimated best fit of the equations of state parameters,  $\omega$CDM and B$\Lambda$CDM are close to $\Lambda$CDM while a smooth deviation is observed for CPL, Barboza-Alcaniz, and logarithmic parameterizations. 

\begin{figure}[t!]
\centering
\begin{minipage}{0.48\textwidth}
    \centering
    \includegraphics[width=\textwidth]{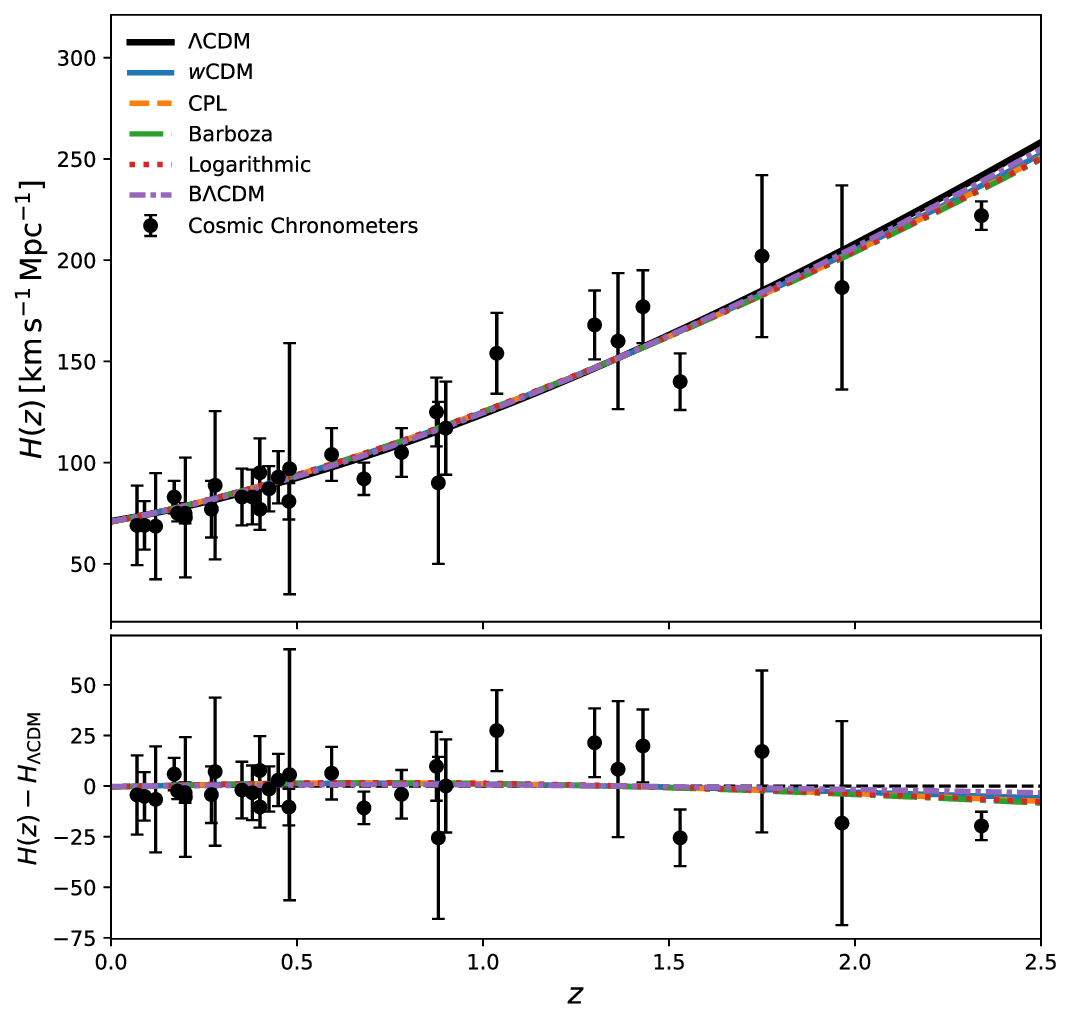}
    \caption{Evolution of the Hubble parameter $H(z)$ (top panel) and the residuals $H(z)-H_{\Lambda\mathrm{CDM}}$ (bottom panel) as functions of redshift $z$, compared with Cosmic Chronometer measurements (black points). The theoretical curves correspond to the $\Lambda$CDM model (black solid line), $w$CDM (blue solid line), CPL (orange dashed line), Barboza-Alcaniz parametrization (green dash-dot line), logarithmic parametrization (red dotted line), and B$\Lambda$CDM (purple dash-dot line).}
    \label{fig:Hubble}
\end{minipage}
\hfill
\begin{minipage}{0.48\textwidth}
    \centering
    \includegraphics[width=\textwidth]{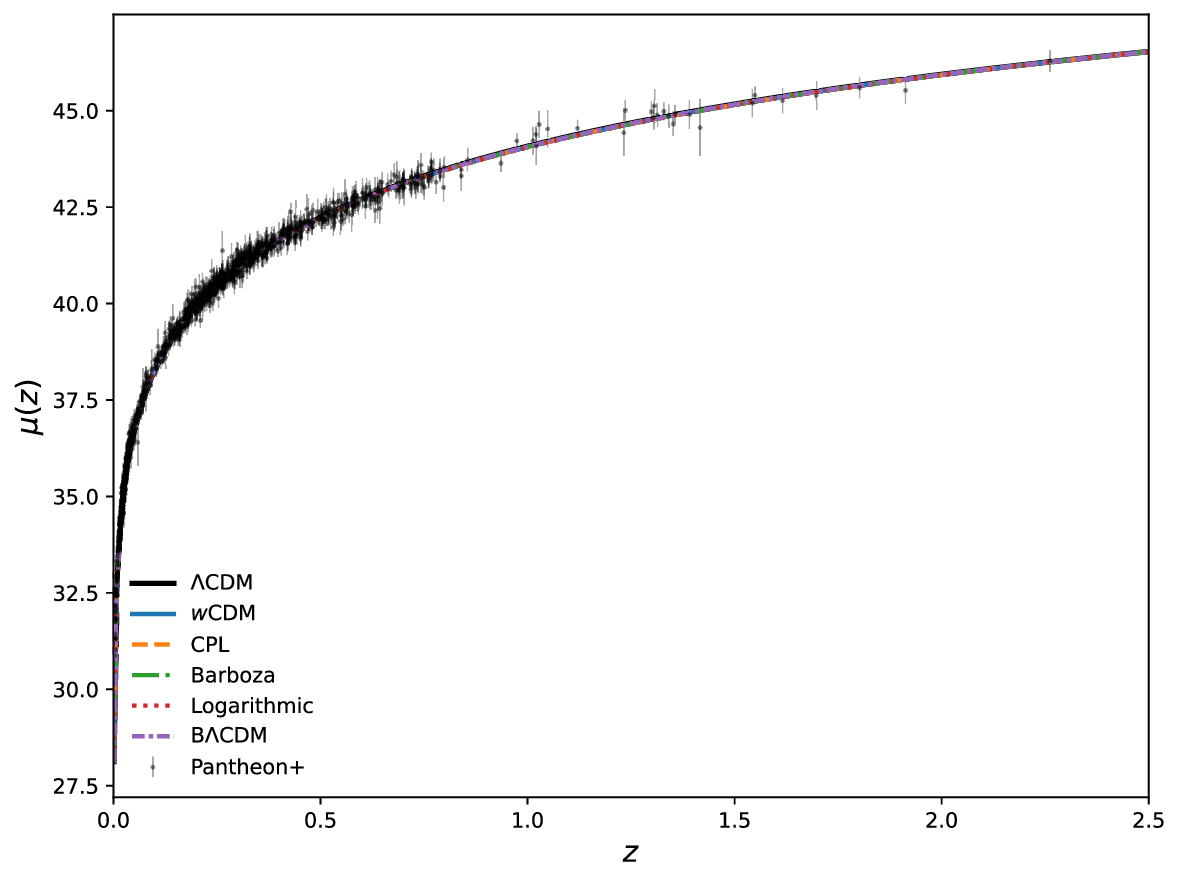}
\caption{Evolution of the predicted distance modulus, $\mu(z)$, as a function of redshift $z$, compared with the Pantheon+ supernova sample (black points). The theoretical curves correspond to the $\Lambda$CDM model (black solid line), $w$CDM (blue solid line), CPL (orange dashed line), Barboza-Alcaniz parametrization (green dash-dot line), logarithmic parametrization (red dotted line), and B$\Lambda$CDM (purple dash-dot line).}
    \label{fig:Pantheon}
\end{minipage}
\end{figure}

\begin{figure}[h!]
\includegraphics[width=0.67\textwidth]{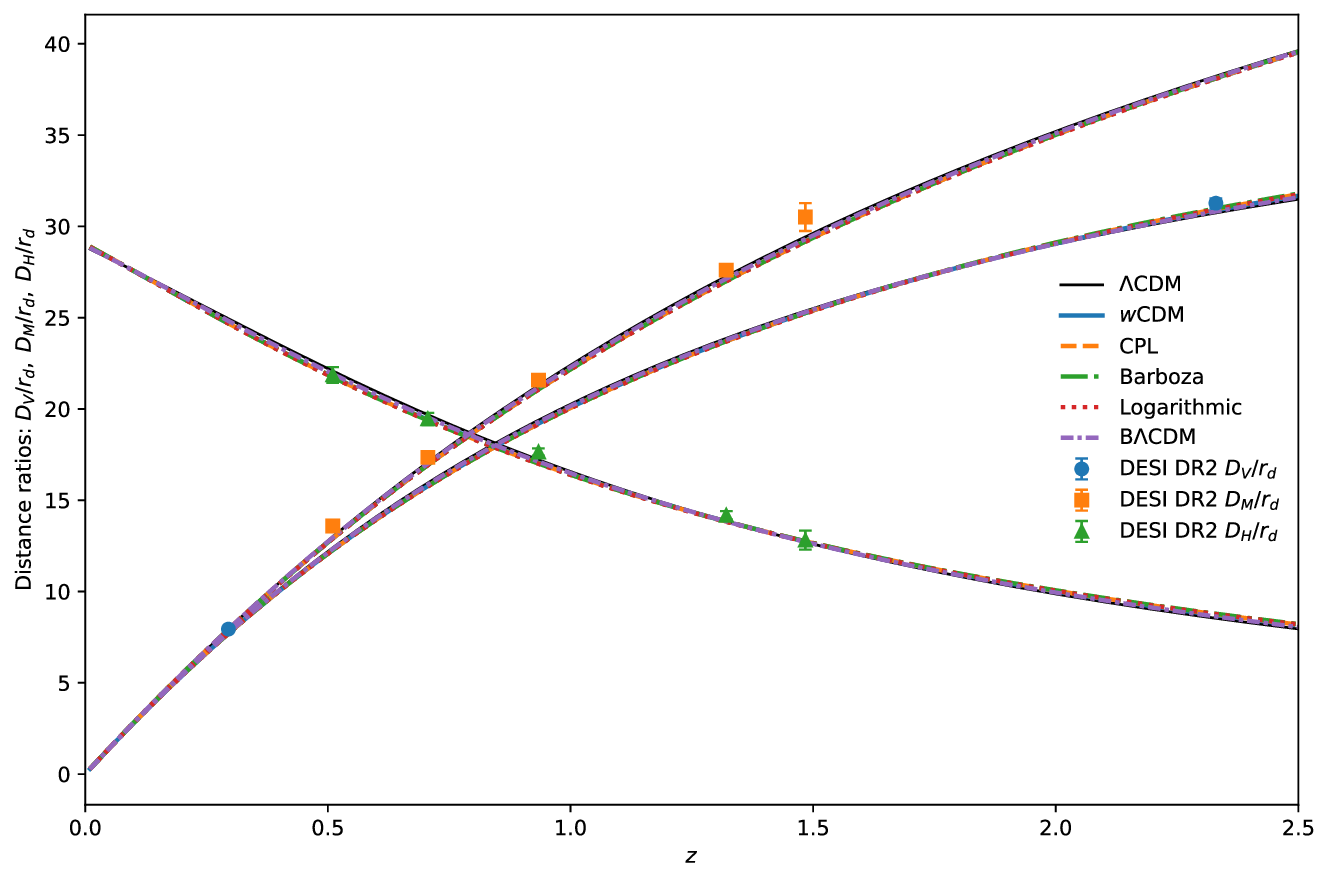}
\caption{Evolution of the distance ratios $D_V/r_d$, $D_M/r_d$, and $D_H/r_d$ as functions of redshift $z$. The DESI DR2 measurements are shown as blue circles ($D_V/r_d$), orange squares ($D_M/r_d$), and green triangles ($D_H/r_d$). The theoretical curves correspond to the $\Lambda$CDM model (black solid lines), $w$CDM (blue solid and dashed lines), CPL (orange dashed lines), Barboza-Alcaniz parametrization (green dash-dot lines), logarithmic parametrization (red dotted lines), and B$\Lambda$CDM (purple dash-dot lines).}
\label{desi}
\end{figure}

In order to compare $\Lambda$CDM and the parameterizations under consideration, the best fit parameters depicted in table \ref{mcmc_models} are used in plotting the variation of the Hubble rate $H(z)$ and the distance modulus $\mu(z)$, Figs. \ref{fig:Hubble} and \ref{fig:Pantheon}, respectively. In addition, Fig.~\ref{desi} presents the evolution of the predicted distance ratios $D_V/r_d$, $D_M/r_d$, and $D_H/r_d$ as functions of $z$, compared with the DESI DR2 measurements. We observe an excellent agreement between the theoretical predictions of all considered models and the observational data, demonstrating their compatibility with current cosmological constraints.\\
\begin{figure}[h!]
\includegraphics[width=0.7\textwidth]{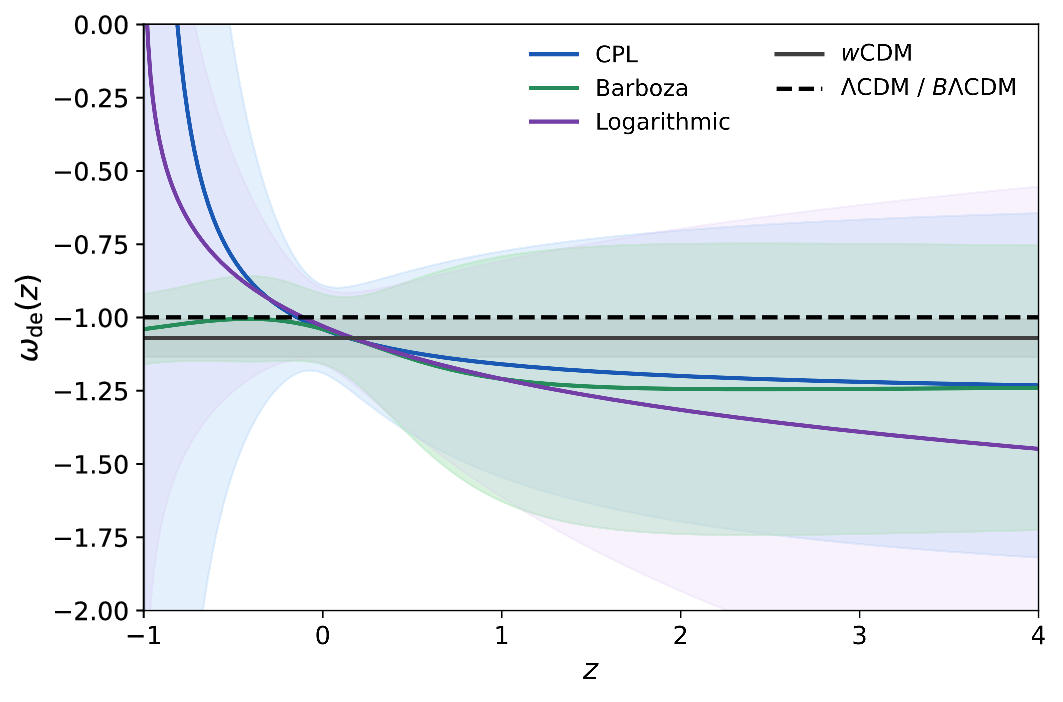}
\caption{Evolution of the equation of state parameter $\omega_{\rm de}(z)$ as a function of redshift $z$. The horizontal black dashed line corresponds to the $\Lambda$CDM model and to the $B\Lambda$CDM case, where $\omega_{\rm de}=-1$ is imposed within the biconnection framework. The solid gray curve represents the $w$CDM model, while the blue, green, and purple solid curves correspond to the CPL, Barboza-Alcaniz, and logarithmic parametrizations, respectively. The shaded regions indicate the $1\sigma$ confidence intervals.}
\label{omega}
\end{figure}
\begin{figure}[h!]
\centering
\begin{minipage}{0.495\textwidth}
    \centering
\includegraphics[width=\textwidth]{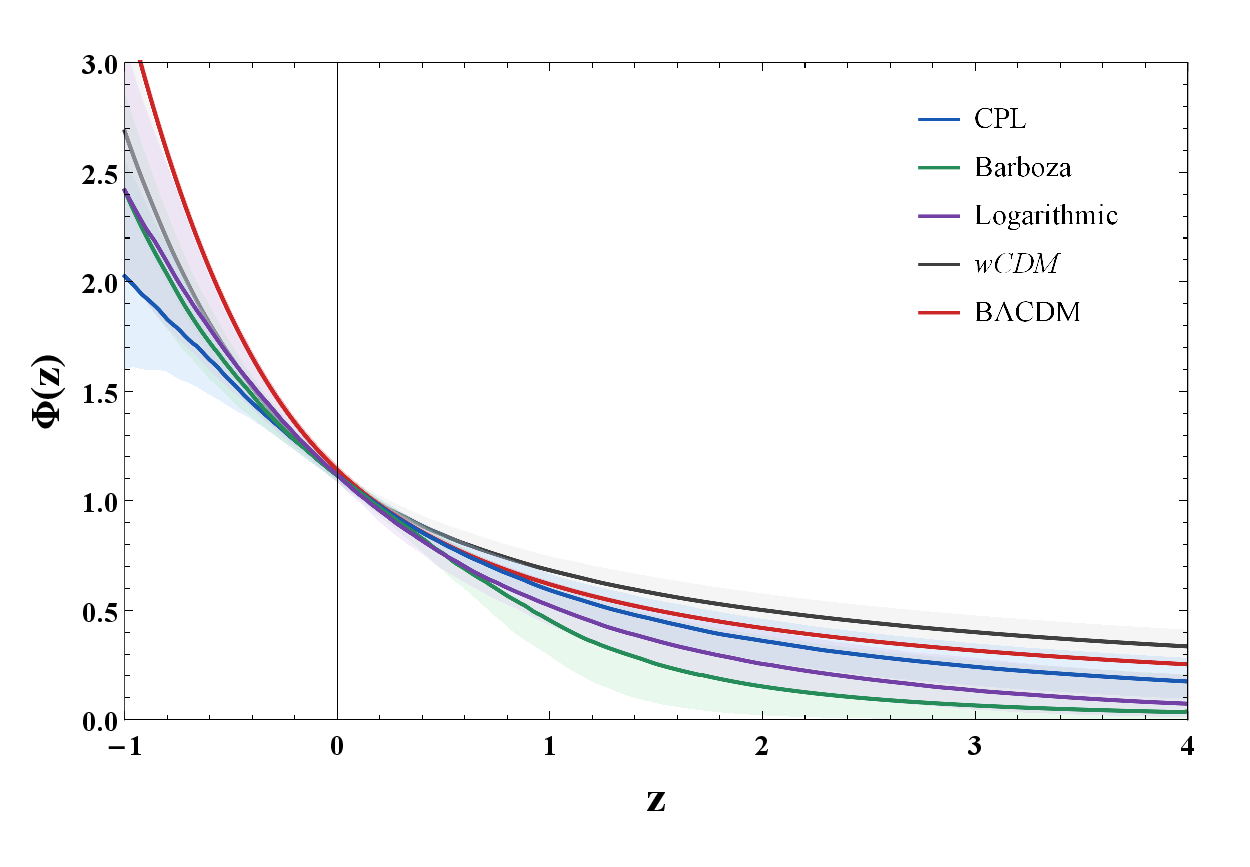}
    \caption{Evolution of the geometrical parameter of the biconnection gravity, $\Phi(z)$, as a function of redshift. The curves correspond to CPL (blue solid), Barboza-Alcaniz (green solid), logarithmic (purple solid), $\omega$CDM (black solid), and $\mathrm{B}\Lambda\mathrm{CDM}$ (red solid). The shaded regions represent the $1\sigma$ confidence intervals.}
    \label{fig:Phi}
\end{minipage}
\hfill
\begin{minipage}{0.495\textwidth}
    \centering
    \includegraphics[width=\textwidth]{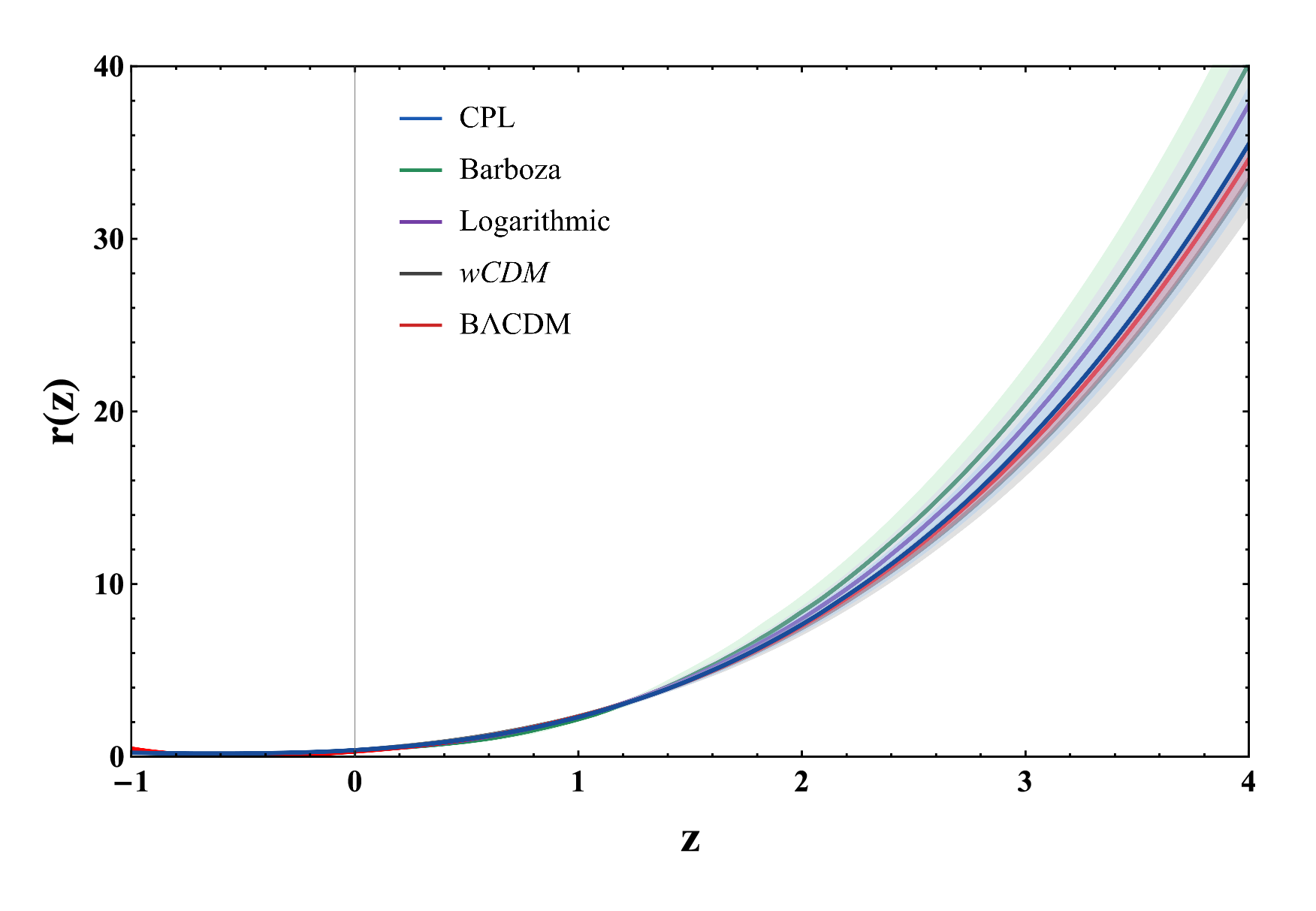}
    \caption{Evolution of the dimensionless matter energy density, $r(z)$, as a function of redshift. The curves correspond to CPL (blue solid), Barboza-Alcaniz (green solid), logarithmic (purple solid), $\omega$CDM (black solid), and $\mathrm{B}\Lambda\mathrm{CDM}$ (red solid). The shaded regions indicate the $1\sigma$ confidence intervals.}
    \label{fig:r}
\end{minipage}
\end{figure}

\begin{figure}[h!]
\includegraphics[width=0.6\textwidth]{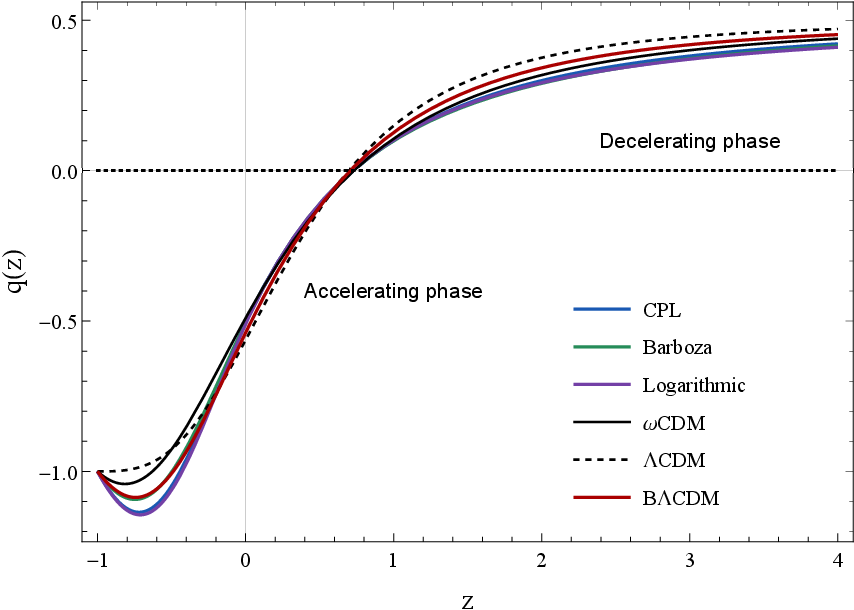}
\caption{Evolution of the deceleration parameter $q(z)$ as a function of redshift. The curves correspond to CPL (blue solid), Barboza-Alcaniz (green solid), logarithmic (purple solid), $\omega$CDM (black solid), $\Lambda$CDM (black dashed), and $\mathrm{B}\Lambda\mathrm{CDM}$ (red solid). The horizontal dotted line marks the transition between accelerating and decelerating phases.}
\label{q}
\end{figure}
\begin{figure}[h!]
\centering
\begin{minipage}{0.48\textwidth}
    \centering
    \includegraphics[width=\textwidth]{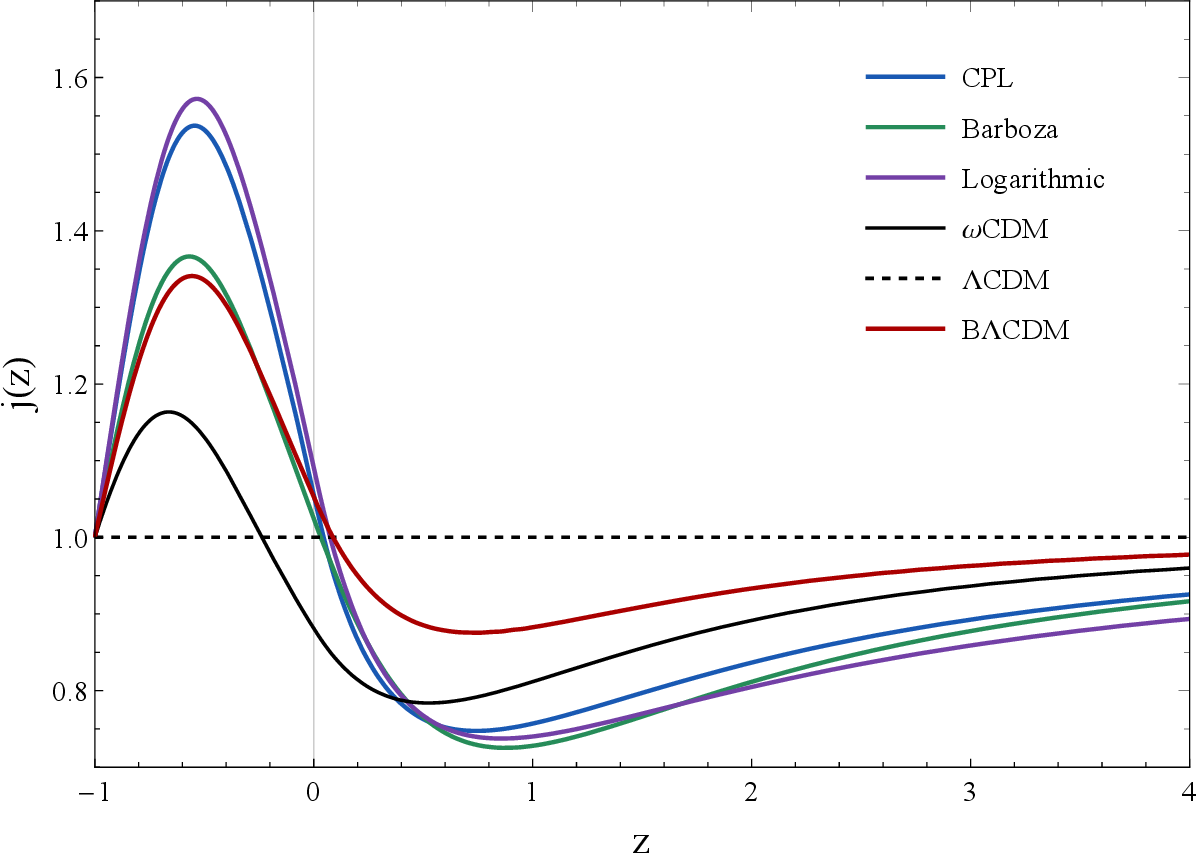}
    \caption{Evolution of the jerk parameter $j(z)$ as a function of redshift. The curves correspond to CPL (blue solid), Barboza-Alcaniz (green solid), logarithmic (purple solid), $\omega$CDM (black solid), $\Lambda$CDM (black dashed), and $\mathrm{B}\Lambda\mathrm{CDM}$ (red solid).}
    \label{fig:jerk}
\end{minipage}
\hfill
\begin{minipage}{0.48\textwidth}
    \centering
    \includegraphics[width=\textwidth]{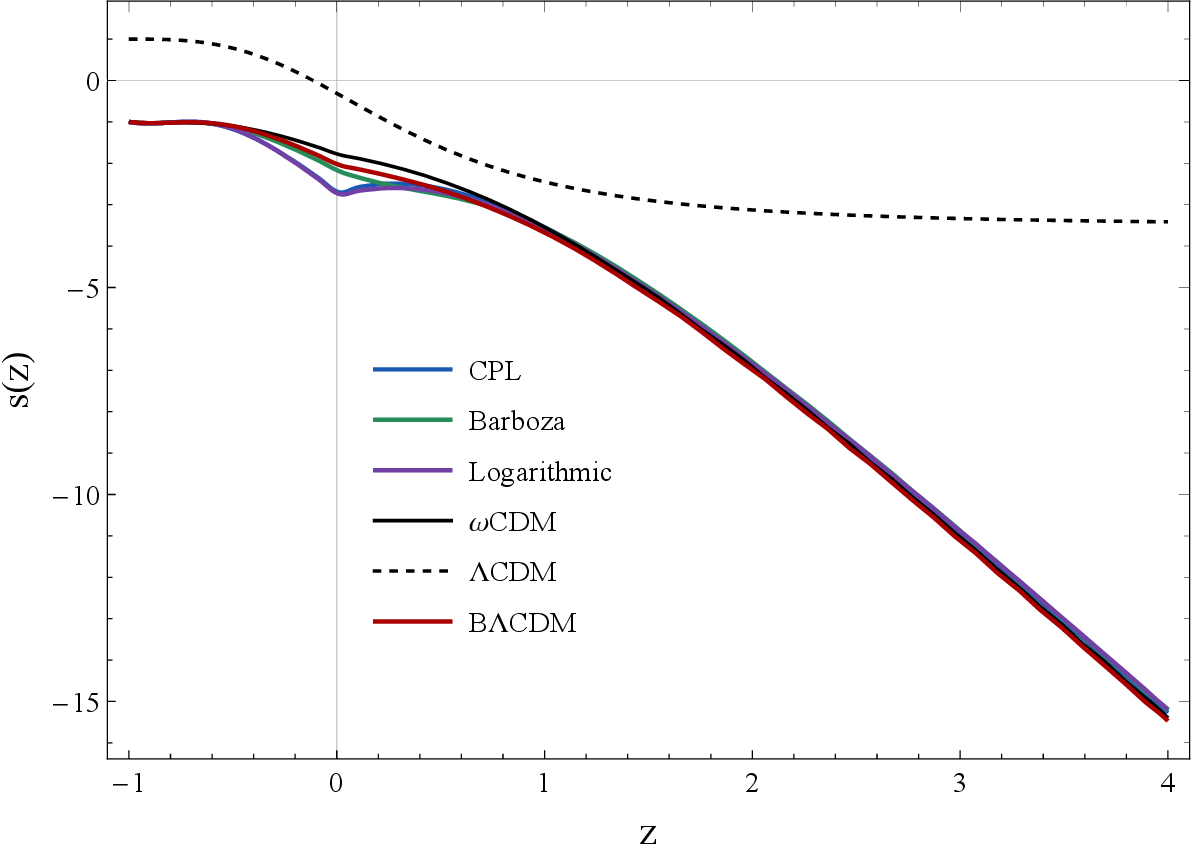}
    \caption{Evolution of the snap parameter $s(z)$ as a function of redshift. The curves correspond to CPL (blue solid), Barboza-Alcaniz (green solid), logarithmic (purple solid), $\omega$CDM (black solid), $\Lambda$CDM (black dashed), and $\mathrm{B}\Lambda\mathrm{CDM}$ (red solid).}
    \label{fig:snap}
\end{minipage}
\end{figure}
\begin{figure}[h!]
\centering
\begin{minipage}{0.49\textwidth}
    \centering
    \includegraphics[width=\textwidth]{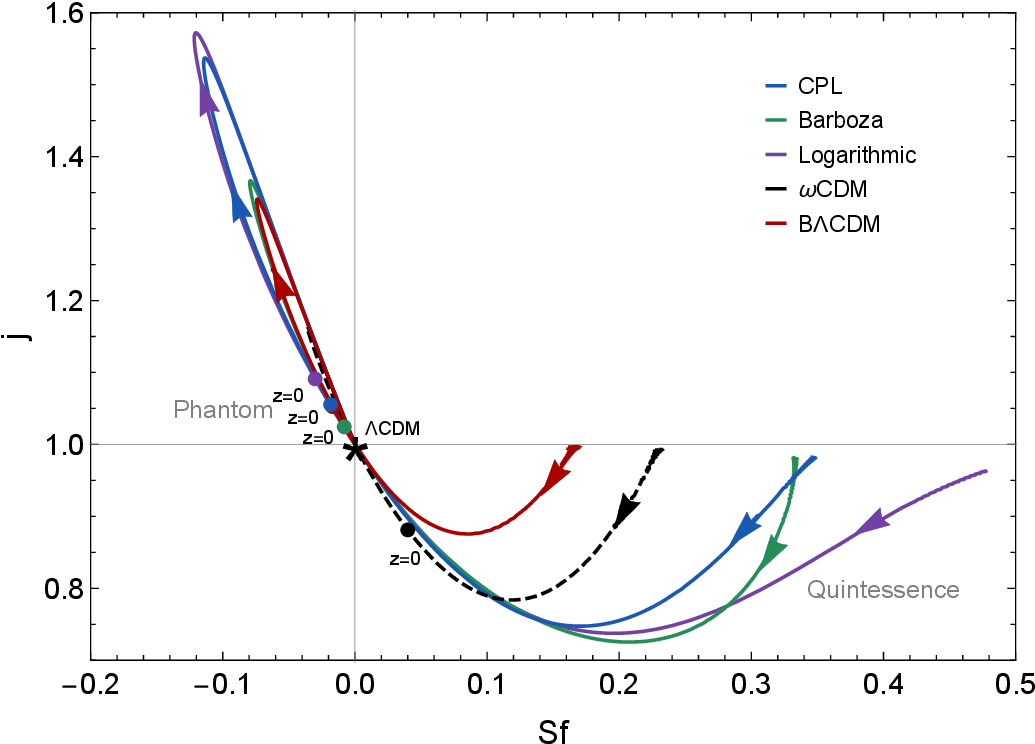}
    \caption{Statefinder plane $(s,j)$ for $\Lambda$CDM (black star), $\omega$CDM (black solid), CPL (blue solid), Barboza-Alcaniz (green solid), logarithmic (purple solid), and $\mathrm{B}\Lambda\mathrm{CDM}$ (red solid). The arrows indicate the dynamical evolution from the past to the future.}
    \label{fig:sj}
\end{minipage}
\hfill
\begin{minipage}{0.49\textwidth}
    \centering
    \includegraphics[width=\textwidth]{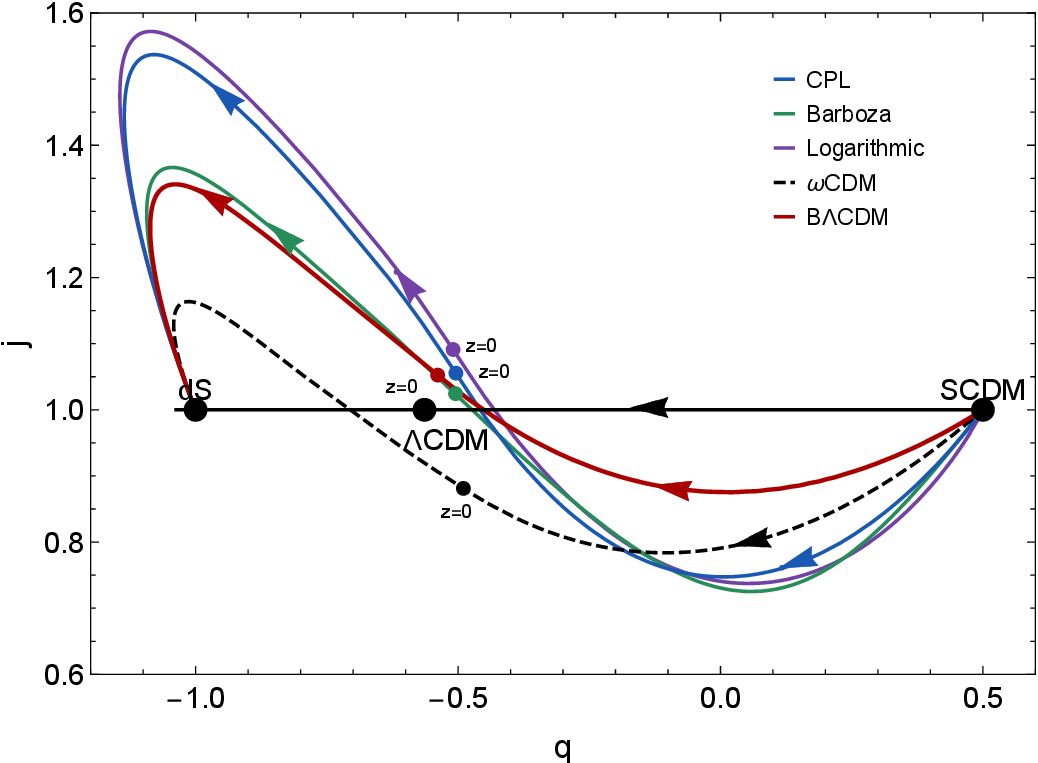}
    \caption{Statefinder plane $(q,j)$ for $\Lambda$CDM (black star), $\omega$CDM (black solid), CPL (blue solid), Barboza-Alcaniz (green solid), logarithmic (purple solid), and $\mathrm{B}\Lambda\mathrm{CDM}$ (red solid). The arrows indicate the dynamical evolution from the past to the future.}
    \label{fig:qj}
\end{minipage}
\end{figure}
\subsection{Biconnection and related parameters }

The parameters related to the biconnection gravity are mainly the EoS parameter defined as the ratio of the effective pressure to the energy density, the torsion vector $\Phi$ and  the matter energy density. As illustrated in Fig. \ref{omega}, $\Lambda$CDM and B$\Lambda$CDM are represented by a dashed horizontal line with $\omega_{de}=-1$, $\omega$CDM plotted as black horizontal line, CPL as a red curve, Barboza-Alcaniz as blue curve, and logarithmic as green curve. $\omega$CDM and Barboza-Alcaniz exhibit an equation of state less than $-1$ i. e. they behave like a phantom. However, CPL and logarithmic equations of state show a phantom behavior in the past, cross the divide line  in the recent past and mimic the quintessence behavior in the future.

 The second parameter characterizing the biconnection gravity, the torsion vector $\Phi$,
is presented in Fig. \ref{fig:Phi}. For
all EoS parameterizations of the biconnection gravity, the torsion vector shows an increasing positive behavior. In the recent past, up to $z=1$, and in the future up to $z\sim -0.25$, all parameterizations are equal in describing the biconnection gravity. The discrepancy between these parameterizations to describe the biconnection gravity appears beyond $z=1$ and below $z\sim -0.25$.

The matter energy density of all parameterizations of the biconnection gravity is shown in Fig. \ref{fig:r}.  As expected, the  matter energy density dominates the budget component of the Universe in the past and it is negligible in the recent past, today and in the future. Furthermore, all parameterizations of the biconnection gravity are at the same footing with respect to the  matter energy density in the recent past.

\subsection{Cosmographic Parameters and Statefinder Diagnostics}

To discriminate among the large number of theoretical models  describing the current phase of cosmic acceleration,  the statefinder diagnostic provides a powerful tool for model comparison. Statefinder diagnostic is based on higher order derivatives of the scale factor, namely the Hubble rate, $H$, the deceleration parameter, $q$, the jerk parameter, $j$, and the snap parameter, $s$. The expression of these parameters are extracted from the Taylor expansion of the scale factor around the present time $t_0$.
\begin{equation}
\begin{gathered}
a(t)=a_0\left[1+\left.\frac{1}{a_0} \frac{d a}{d t}\right|_{t_0}\left(t-t_0\right)+\left.\frac{1}{2 !a_0} \frac{d^2 a}{d t^2}\right|_{t_0}\left(t-t_0\right)^2\right. \\
\left.+\left.\frac{1}{3 !a_0} \frac{d^3 a}{d t^3}\right|_{t_0}\left(t-t_0\right)^3+\left.\frac{1}{4 ! a_0} \frac{d^4 a}{d t^4}\right|_{t_0}\left(t-t_0\right)^4+\mathcal{O}\left(t-t_0\right)^5\right],
\end{gathered}
\label{a}
\end{equation}
where the current value of the scale factor is denoted by $a_0$. From the coefficients of the Taylor series expansion, the expressions of the Hubble parameter,  deceleration, jerk  and snap parameters are
\begin{equation}
\begin{gathered}
H=\frac{1}{a} \frac{d a}{d t}, \hspace{2cm} q=-\frac{1}{a H^2} \frac{d^2 a}{d t^2}, \\ 
 j=\frac{1}{a H^3} \frac{d^3 a}{d t^3}, \hspace{2cm} s=\frac{1}{a H^4} \frac{d^4 a}{d t^4},
\end{gathered}
\end{equation}
respectively.
\bigskip
 To simplify our analysis, we express the jerk and the snap  parameters in terms of the
redshift and the deceleration parameter. These parameters transformation writes as 
\begin{eqnarray}
q&=&-1+ (1+z)\frac{1}{H(z)}\frac{dH(z)}{d z},\nonumber\\ 
j&=& (1+z) \frac{d q}{d z}+q(2 q+1), 
\end{eqnarray}
and 
\begin{eqnarray}
s&=& -(1+z) \frac{d j}{d z}-j(2+3 q).\nonumber
\end{eqnarray}

\paragraph{The deceleration parameter.} The cosmic acceleration of the expansion of the Universe is quantified by the deceleration parameter, $q(z)$, whose evolution is shown in Fig.~\ref{q}. 
All biconnection models exhibit a transition from a decelerated expansion phase in the past to an accelerated phase at late times. 
The transition redshift is found to be 
$z_t = 0.695^{+0.023}_{-0.023}$, 
$0.705^{+0.024}_{-0.024}$, 
$0.73 \pm 0.03$, 
$0.72 \pm 0.05$, 
$0.72^{+0.05}_{-0.05}$, 
and $0.73^{+0.06}_{-0.06}$ 
for $\Lambda$CDM, $B\Lambda$CDM, $w$CDM, CPL, Barboza--Alcaniz, and logarithmic parameterizations, respectively. 
These values are statistically consistent within $1\sigma$, indicating that all models predict a similar onset of the present accelerated phase.\\

\paragraph{The jerk parameter.} The evolution of the jerk parameter, $j(z)$, with respect to the redshift is shown in Fig. \ref{fig:jerk}.  We notice from this figure that the behavior of the jerk parameter is similar for $\omega$CDM, and for CPL, BA and Logarithmic parameterizations. Indeed, compared to $\Lambda$CDM, these parameterizations  decrease from the pass and begin to increase from the recent pass to the near future and observe a decreasing behavior once more. Furthermore, except the $\omega$CDM case, the present values of the jerk parameter of all parameterizations  are a little bit greater than the one  of $\Lambda$CDM. These behaviors indicate that currently the length preserving biconnection gravity mimics a phantom trend. \\

\paragraph{The snap parameter.} The evolution of the snap parameter versus the redshift of each parameterizations is shown in 
Fig.  \ref{fig:snap}. All parameterizations as well as the $\Lambda$CDM model exhibit negative values of the snap parameter. Furthermore, this figure shows a very similar  behavior of all parameterizations in the past i.e. no preference of the length preserving biconnection gravity regarding these parameterizations. However a discrimination between these parameterizations began to appear in the recent past and in the near future. \\

\paragraph{The statefinder pairs.} To gain more insight into the expansion dynamics, we investigate the statefinder parameter pair $\{j,s_f\}$, which has been extensively discussed in the literature as a powerful diagnostic for dark energy models \cite{sr1,gor2003,sr3}. This construction parameters provide a more detailed characterization of cosmic evolution than the Hubble parameter and the deceleration parameter. In particular, the 
$\Lambda$CDM model corresponds to a fixed point, namely $(s_f=0,j=1)$, which serves as a useful benchmark for comparing alternative dark energy scenarios such as quintessence $(s_f>0,j<1)$, Chaplygin gas or phantom $(s_f<0,j>1)$, and standard cold dark matter (SCDM) $(s_f=1,j=1)$. The expression of the statefinder parameters is constructed by setting
\begin{align}
j &=(1+z) \frac{d q}{d z}+q(2 q+1),\\
s_f&=\frac{j-1}{3\left(q-\frac{1}{2}\right)}.
\end{align}
The evolution  of the statefinder diagnostic pair $\{s_f, j\}$  is shown in Fig. \ref{fig:sj}.  We observe a similarities in theirs dynamical evolutions. Indeed, All parameterizations progress from the past within the region of $s_f > 0$ and $j < 1$ which means an evolution from the quintessence dynamics to the phantom dynamics in the future  corresponding to the region of $s_f<0$ and $j>1$. We notice that all parameterizations cross the $\Lambda$CDM model corresponding to $s_f=0$ and $j=1$ but at different redshifts. Indeed, B$\Lambda$CDM, $\omega$CDM, CPL, Barboza-Alcaniz and logarithmic parameterizations cross $\Lambda$CDM at $z=0.0879$, $z=-0.237$, $z=0.0452$, $z=0.0316$, and $z=0.0751$, respectively. At these redshifts, the biconnection gravity behaves like $\Lambda$CDM.
\\

From the evolutionary trajectory in the statefinder $j-q$ plane, the current point $(q_0=-0.564, j_0=1)$ corresponds to the $\Lambda$CDM model and the point $(q_0=0.5, j_0=1)$ corresponds to the SCDM model while the point $(q_0=-1, j_0=1)$ is consistent with de Sitter expansion. In our setup,  the biconnection gravity corresponds to the current  point $(q_0=-0.54, j_0=0.22)$, $(q_0=-0.49, j_0=0.16)$, $(q_0=-0.52, j_0=0.47)$, $(q_0=-0.52, j_0=0.40)$, and $(q_0=-0.52, j_0=0.47)$ for B$\Lambda$CDM, $\omega$CDM, CPL, Barboza-Alcaniz and logarithmic parameterizations, respectively.
From Fig. \ref{fig:qj}, we observe that the behavior of the statefinder $\{q, j\}$ of biconnection parameterizations starts from the point of SCDM in the past and evolve to the de Sitter point in the future. These parameterizations evolve from the decelerating region, $q >0$ and $j<1$, to the accelerating region, $q<0$ and $j>1$. We notice also that from the recent past to the future, B$\Lambda$CDM, CPL, Barboza-Alcaniz and logarithmic parameterizations describe the accelerating phase.
\subsection{ $Om(z)$ diagnostic }\label{E}
\begin{figure}[h!]
\includegraphics[width=0.6\textwidth]{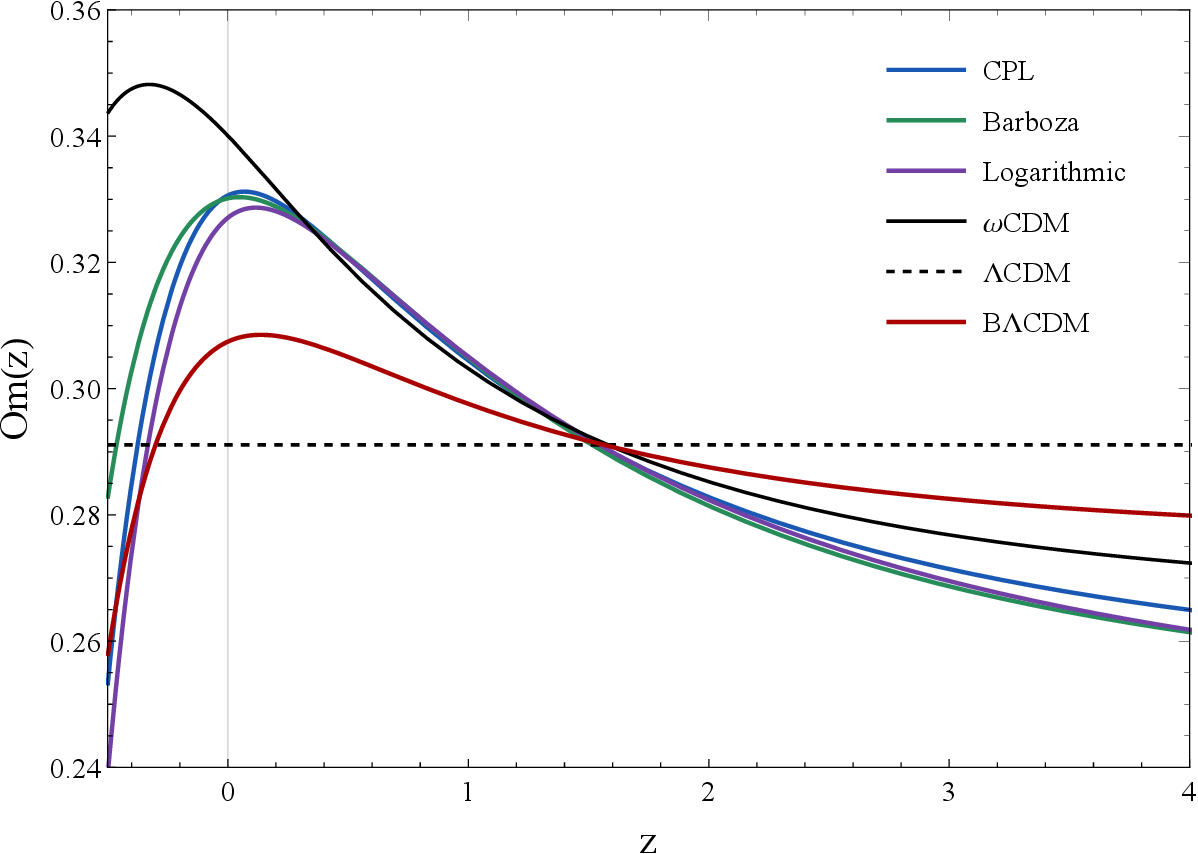}
\caption{Evolution of the $Om(z)$ diagnostic as a function of redshift. The curves correspond to CPL (blue solid), Barboza-Alcaniz (green solid), logarithmic (purple solid), $\omega$CDM (black solid), $\Lambda$CDM (black dashed), and $\mathrm{B}\Lambda\mathrm{CDM}$ (red solid).}
\label{Om}
\end{figure}

The $Om(z)$ diagnostic and the statefinder parameters serve as complementary probes of dark energy dynamics. The $Om(z)$  test provides a purely geometrical criterion that distinguishes between  cosmological constant and dynamical dark energy models without requiring precise knowledge of the present matter density parameter $\Omega_{m0}$ \cite{Om2,Om3}. By contrast, the statefinder pair $(s,j)$, constructed from higher order derivatives of the scale factor, encodes additional information about the expansion history. Models exhibiting  weak redshift dependence in $Om(z)$  are typically associated with trajectories that remain close to the $\Lambda$CDM
where $(s,j)=(0,1)$, whereas pronounced variations in $Om(z)$  correspond to significant departures from this point in the statefinder plane. Consequently, the joint analysis of these diagnostics provides a robust and internally consistent framework for discriminating among competing dark energy models.
  On the other hand, a positive slope of $Om(z)$ quantifies a phantom behavior of dark energy
whereas negative slope implies that dark energy behaves like quintessence.  For a flat Universe, the $Om(z)$ diagnostic is defined as
\begin{equation}
O m(z) \equiv \frac{h^2(z)-1}{(1+z)^3-1},
\end{equation}
In Fig. \ref{Om}, we illustrate the evolution of $Om(z)$ diagnostic as a function of the redshift for $\Lambda$CDM model and for the other parameterizations of the biconnection gravity.  All parameterizations cross the $\Lambda$CDM point at almost the same redshift.  From this figure, Barboza-Alcaniz and logarithmic behave similarly while a clear discrepancy  begin to appear from the recent past to the future.  Fig. \ref{Om} shows negative slope for all parameterizations. This means that the biconnection gravity behaves like quintessence. Except $\omega$CDM, the parameterizations B$\Lambda$CDM, CPL, Barboza-Alcaniz and logarithmic exhibits positive slope from the recent past to the  future where the biconnection gravity behaves like phantom.

\section{Conclusions}\label{sec6} 

In this work, we have investigated the cosmological implications of a length-preserving biconnection gravity constructed from the Schrödinger connection and its dual. In this framework, the symmetric combination of the two connections reduces to the Levi--Civita connection, thereby recovering General Relativity at the background level, while their difference encodes non-Riemannian geometric degrees of freedom through the mutual curvature tensor. The resulting modified Friedmann equations contain additional geometric contributions that naturally behave as an effective dark energy component.\\

To assess the observational viability of this framework, we parameterized the effective dark energy sector using five equations of state: B$\Lambda$CDM, $\omega$CDM, Chevallier--Polarski--Linder (CPL), Barboza--Alcaniz, and logarithmic parameterizations. The cosmological parameters were constrained using a Markov Chain Monte Carlo analysis based on the combined DESI DR2 + Pantheon+ + Cosmic Chronometer dataset.\\

All parameterizations provide an excellent fit to current observational data, with reduced chi-square values close to unity. The present value of the Hubble parameter is remarkably stable across all models, $H_0 \approx 70.7\,\mathrm{km\,s^{-1}\,Mpc^{-1}}$, indicating that the biconnection framework does not significantly modify the late-time expansion rate. The equation-of-state parameters are statistically consistent with $w_{de} = -1$. The torsion-related parameter $\Phi_0$ is consistently constrained around $\Phi_0 \simeq 1.11$, showing stability across different parameterizations.\\

A statistical comparison using information criteria provides a clear classification of the models. According to AICc and DIC, the ranking is $\text{B}\Lambda\text{CDM},\omega\text{CDM}, \text{CPL}, \text{Logarithmic}, \text{Barboza--Alcaniz}, \Lambda\text{CDM}$. In contrast, the BIC ranking is $\text{B}\Lambda\text{CDM}, \Lambda\text{CDM} , \omega\text{CDM},  \text{CPL}, \text{Logarithmic}, \text{Barboza--Alcaniz}$. The difference arises from the stronger penalization of additional parameters in BIC. While AICc and DIC reward improved fit quality, BIC favors simpler models. Nevertheless, B$\Lambda$CDM consistently emerges as the most supported model across all criteria, indicating that the geometric extension introduced by the biconnection framework is statistically competitive with $\Lambda$CDM.\\

The cosmographic analysis further clarifies the physical implications of the model. All parameterizations predict a transition from a decelerated to an accelerated phase at $z_t \sim 0.7$, in excellent agreement with $\Lambda$CDM. The present deceleration parameter $q_0 \approx -0.5$ confirms the accelerated expansion. A distinctive feature emerges in the evolution of the jerk parameter. A particularly interesting case is the B$\Lambda$CDM model, for which the effective equation of state satisfies $w_{\rm de}=-1$ (i.e., $p_{\rm eff}=-\rho_{\rm eff}$), making it formally equivalent to $\Lambda$CDM in terms of the EoS. However, unlike $\Lambda$CDM, the B$\Lambda$CDM model shows a non-constant jerk parameter, reflecting the modified background evolution introduced by the biconnection geometric framework. This indicates that even though the effective dark energy behaves like a cosmological constant in terms of its pressure, the expansion dynamics is intrinsically non-$\Lambda$CDM. The biconnection models considered in the present work show a dynamical behavior in which the effective evolution initially corresponds to a quintessence-like regime and subsequently evolves toward a phantom-like regime. All models face this transition in the recent past, just before $z=0$, except for the $w$CDM model, which enters the phantom-like regime only in the future. This transition reflects a crossing of the 
phantom divide, indicating that the effective dark energy component generated by the non-Riemannian geometric degrees of freedom is intrinsically dynamical. Such behavior indicates a richer structure than a simple cosmological constant, 
while remaining compatible with observational constraints.\\

The Om$(z)$ diagnostic reinforces these results. While $\Lambda$CDM predicts a 
constant Om$(z)$, the B$\Lambda$CDM model shows the closest behavior to this 
limit. The $\omega$CDM, CPL, Logarithmic, and Barboza--Alcaniz models 
depict a more noticeable evolution driven by their time-dependent equation 
of state. Overall, Om$(z)$ confirms that the geometric sector generates a 
 dynamical dark energy component that remains observationally close 
to $\Lambda$CDM at the background level.\\

It is important to mention that the present analysis is restricted to the background level. Since the symmetric part of the biconnection construction reproduces General Relativity at the homogeneous level, deviations from $\Lambda$CDM manifest effectively. A more stringent test of the theory therefore requires a full perturbative treatment and confrontation with structure formation data, including redshift-space distortions and CMB observations. The inclusion of early-Universe probes such as Planck data will be crucial for assessing the viability of the model across all cosmological epochs.\\

In conclusion, the length-preserving biconnection gravity provides a geometrically well-motivated extension of General Relativity in which cosmic acceleration emerges from non-Riemannian geometric degrees of freedom. Current late-time observational data show that this framework is statistically competitive with $\Lambda$CDM and dynamically consistent with the observed expansion history. Future studies at the perturbative level will provide a deeper understanding of the  role played by the non-Riemannian geometric degrees of freedom in the cosmic evolution.

\end{document}